\documentclass[aps,twocolumn,prb,showpacs,amsmath,amssymb,citeautoscript,nobibnotes]{revtex4}
\makeatletter
\def\@dotsep{4.5}
\makeatother

\usepackage{graphicx}

\newcommand{\rve}{\boldsymbol r}

\newcommand{\eve}{\boldsymbol e}
\newcommand{\bem}{\begin{em}}
\newcommand{\eem}{\end{em}}
\newcommand{\bq}{\begin{eqnarray}}
\newcommand{\eq}{\end{eqnarray}}
\newcommand{\bqn}{\begin{eqnarray*}}
\newcommand{\eqn}{\end{eqnarray*}}

\newcommand{\beq}{\begin{equation}}
\newcommand{\eeq}{\end{equation}}

\newcommand{\der}[2]{\frac{d #1}{d #2}}    
    
\newcommand{\etal}{\bem et al.\eem}    

\begin{document}
  
\title{Understanding fragility in supercooled Lennard-Jones
  mixtures.\\ II. Potential energy surface} 
\author{D. Coslovich}
\email{coslo@.ts.infn.it}
\author{G. Pastore}
\email{pastore@.ts.infn.it}
\affiliation{Dipartimento di Fisica Teorica, Universit{\`a} di Trieste
  -- Strada Costiera 11, 34100 Trieste, Italy} 
\affiliation{CNR-INFM Democritos National Simulation Center --
  Via Beirut 2-4, 34014 Trieste, Italy} 
\date{\today}
\begin{abstract}
We numerically investigated the connection between isobaric fragility
and the properties of high-order stationary points of the potential energy
surface in different supercooled Lennard-Jones mixtures. The increase of
effective activation energies upon supercooling appears
to be driven by the increase of average potential energy barriers
measured by the energy dependence of the fraction of unstable modes.
Such an increase is sharper, the more fragile is the mixture. Correlations
between fragility and other properties of high-order stationary
points, including the vibrational density of states and 
the localization features of unstable modes, are also discussed.    
\end{abstract}
\pacs{61.43.Fs, 61.20.Lc, 64.70.Pf, 61.20.Ja}

\maketitle

\section{Introduction}

The study of the Potential Energy Surface (PES), or energy
landscape, of supercooled liquids and glasses is of
fundamental importance for understanding  
thermodynamical and dynamical
properties in these systems~\cite{sciortino05,book:wales}. Since the
pioneering work of Stillinger and Weber~\cite{sw82},  
a growing body of data, coming from numerical simulations,
has provided a detailed description of the energy landscape explored
by supercooled liquids. As the temperature is lowered toward the
glass transition, progressively deeper regions of the PES
are visited \cite{sastry98}, where the basins of attraction of groups
of local minima (metabasins~\cite{stillinger}) act as traps for the
system in configuration space~\cite{denny03,doliwa03,doliwa03a,doliwa03b} and
slow down the liquid dynamics. Accordingly, viscosity and structural
relaxation times show a dramatic increase upon supercooling. In the so
called fragile glass-formers, such an increase is faster than
Arrhenius (or super-Arrhenius), as opposed to the behavior of strong
glass-formers, in which the temperature dependence of transport
coefficients roughly follows the Arrhenius law.

Schematic descriptions of the PES have often been invoked to 
explain the fragile versus strong behavior of supercooled
liquids~\cite{stillinger95}. Strong glass-formers are expected to have
a rough energy landscape, with energy barriers whose amplitude is essentially
independent of the energy level. On the other hand, fragile
glass-formers should display a more complex organization of stationary
points and a broader distribution of energy barriers. Understanding, at a
quantitative level, the varying degree of fragility in different
glass-formers represents a formidable task for 
theories. Correlations between fragility and statistical or
vibrational properties of local minima of the
PES~\cite{speedy99,sastry01} have recently received 
a critical assessment for a wide range of models of the
PES~\cite{ruocco04}. Variations in the properties of the PES
explored by supercooled liquids at different densities have also been
discussed~\cite{sastry01}, but their correlation to fragility have
been a matter of debate~\cite{tarjus04}. Detailed studies
have focused on the 
role of elementary rearrangements between adjacent local minima
through transition states, both at constant
density~\cite{middleton01b} and constant pressure~\cite{middleton03}.
Here we will concentrate our attention on the properties of high-order
stationary points of the PES, whose 
relevance for supercooled liquids has been emphasized 
in the last
years~\cite{angelani00,angelani02,doye02,wales03,angelani03,sampoli03,coslovich06}.
High-order stationary points could 
offer a simple explanation of the fragile behavior of glass-formers,
in terms of an increase of average energy
barriers~\cite{cavagna01c}. This feature is encoded, in an effective
way, in a number of models of energy landscapes developed in the last
years~\cite{cavagna01b,zamponi03,keyes02b,andronico04}, and has 
sometimes been addressed in numerical simulations~\cite{grigera02}.

Statistical properties of high-order stationary points of the PES
have been investigated recently for a variety of monoatomic and binary
systems, both in the liquid~\cite{shah01,chakraborty06} and
supercooled regime~\cite{angelani03,angelani04}.
The existence of some universal features in the energy landscape of
different model liquids~\cite{angelani03,angelani04} has been
highlighted. At least in the case of the modified soft-sphere
mixtures studied in Ref.~\onlinecite{angelani04}, such a universality has been found to
reflect a fragility invariance of the systems
investigated~\cite{demichele04}. In this work, we take a complementary
point of view and ask: Are there \bem variations \eem in the
properties of high-order stationary points which correlate to fragility? 
To address this point, we consider a set of Lennard-Jones mixtures
cooled at constant pressure: a series of 
equimolar, additive mixtures with varying size ratio, together with
some well-studied binary mixtures
(Sec.~\ref{sec:model}). By investigating the temperature 
dependence of effective activation energies for
relaxation~(Sec.~\ref{sec:fragility}), we provide
support to our previous results~\cite{coslovich07a}, which indicated the
existence of systematic variations of isobaric fragility in additive
mixtures and a remarkable pressure invariance in the mixture of Kob and
Andersen~\cite{ka1,ka2}. These trends allow us 
to test the connection between fragility and some statistical
properties of stationary points of the PES~(Sec.~\ref{sec:pes}). In
particular, we show how fragility can be reflected in the saddles'
density of states, average energy barriers and localization properties
of unstable modes.  

\section{Models and simulation techniques}\label{sec:model} 

The binary mixtures studied in this work consist of $N$
classical particles interacting via the Lennard-Jones potential
\beq \label{eqn:lj}
u_{\alpha\beta}(r) = 4 \epsilon_{\alpha\beta} \left[ {\left( 
    \frac{\sigma_{\alpha\beta}}{r} \right)}^{12} -
  {\left( \frac{\sigma_{\alpha\beta}}{r} \right)}^6 \right]
\eeq
where $\alpha,\beta=1,2$ are indexes of species. Each system is
enclosed in a cubic box with periodic boundary conditions. 
In the following, reduced Lennard-Jones units will be used, 
i.e., $\sigma_{11}$, $\epsilon_{11}$ and
$\sqrt{m_1\sigma_{11}^2/\epsilon_{11}}$ as units of
distance, energy and time, respectively. Most
simulations have been performed for samples of $N=500$ particles,
employing the cutoff scheme of Stoddard and Ford~\cite{stoddard73} at
a cutoff radius $r_c=2.5$. This cutoff scheme (QS) adds a shift
and a quadratic term to the potential in order to ensure continuity up 
to the first derivative of $u_{\alpha\beta}$ at $r=r_c$. The role of the
continuity of derivatives at the cutoff radius $r_c$ has been
discussed in connection to minimization procedures~\cite{shah03}. To
further investigate this point, we have also tested cut-and-shifted (CS) and
cubic-splined cutoff (CSPL)~\cite{grigera02} on smaller samples
composed by $N=108$ particles. In this case, a slightly smaller value
of the cutoff radius has been used ($r_c=2.2$).   

\begin{table}
\caption{\label{tab:sim}
  Summary of thermal histories and simulation details. Also shown are
  the number concentration of large particles $x_1$, 
  the cutoff scheme used (see text for definitions) and the value of
  the cutoff radius $r_c$. In the case of 
  AMLJ-$\lambda$ mixtures, the following values of
  $\lambda$ have been considered: $\lambda=0.60,0.64,0.70,0.73,0.76,0.82$.} 
\begin{ruledtabular}
\begin{tabular}{llllllll}
     & \multicolumn{7}{c}{Isobaric quenches} \\
\cline{2-8}
     & $N$ &$x_1$& $P$  & Cut-off & $r_c$ \\    
\hline
BMLJ & 108 & 0.8 & 10 & QS      & 2.2        \\ 
     & 108 & 0.8 & 10 & CS      & 2.2        \\ 
     & 108 & 0.8 & 10 & CSPL    & 2.2        \\ 
     & 500 & 0.8 &  5 & QS      & 2.5        \\ 
     & 500 & 0.8 & 10 & QS      & 2.5        \\ 
     & 500 & 0.8 & 20 & QS      & 2.5        \\ 
     & 500 & 0.8 & 50 & QS      & 2.5        \\ 
\hline
WAHN & 108 & 0.5 & 10 & QS      & 2.2        \\ 
     & 500 & 0.5 & 10 & QS      & 2.5        \\ 
     & 500 & 0.5 & 20 & QS      & 2.5        \\ 
\hline
AMLJ-$\lambda$ & 500 & 0.5& 10 & QS      & 2.5        \\ 
\hline
     & \multicolumn{7}{c}{Isochoric quenches} \\
\cline{2-8}
     & $N$ &$x_1$& $\rho$ & Cut-off & $r_c$ \\    
\hline
BMLJ & 500 & 0.8 & 1.2  & QS      & 2.5        \\ 
WAHN & 500 & 0.5 & 1.3  & QS      & 2.5        \\ 
\end{tabular}

\end{ruledtabular}
\end{table}

We will focus our attention on the following binary Lennard-Jones
mixtures: (i) The BMLJ mixture of Kob and Andersen~\cite{ka1},
probably the most widely employed model for numerical simulations of
the glass-transition. (ii) The WAHN mixture of
Wahnstr\"om~\cite{wahnstrom}, which is another well-studied model glass-former.
(iii) A set of additive, equimolar mixtures called AMLJ-$\lambda$,
characterized by different values of size ratio
$\lambda=\sigma_{22}/\sigma_{11}$. In this case, the size ratio 
is varied in the range $0.60\leq\lambda\leq 0.82$. A summary of all
interaction parameters, together with a more detailed
description of these models, can be found in Ref.~\onlinecite{coslovich07a}.

\begin{figure}
\includegraphics*[width=0.46\textwidth]{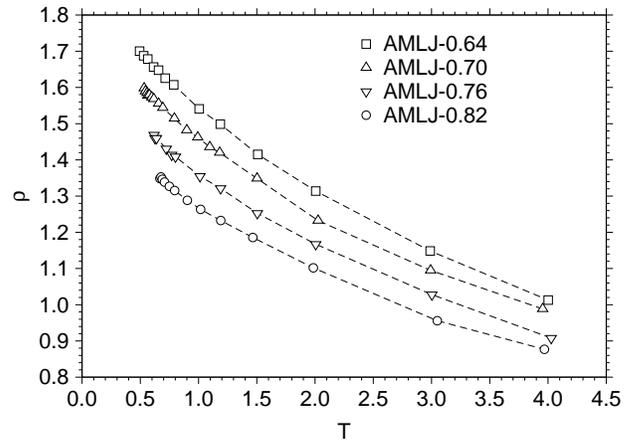}
\caption{\label{fig:presrho1}
  Temperature dependence of density $\rho(T)$ along isobaric
  quenches at $P=10$ for a selection of AMLJ-$\lambda$ mixtures. From
  bottom to top: $\lambda=0.82,0.76,0.70,0.64$.}
\end{figure}

\begin{figure}
\includegraphics*[width=0.46\textwidth]{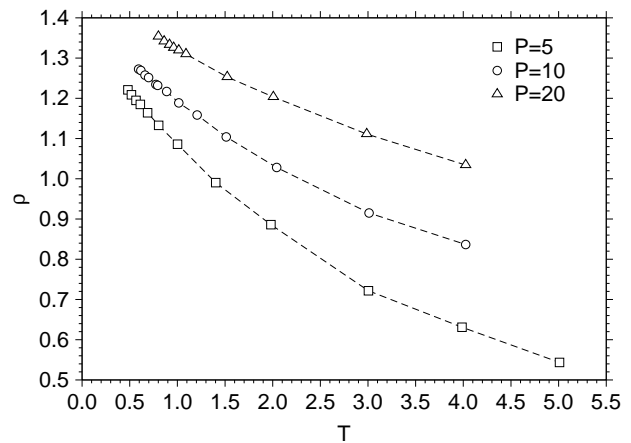}
\caption{\label{fig:presrho2}
  Temperature dependence of density $\rho(T)$ along isobaric
  quenches at for BMLJ at $P=5$, $P=10$ and $P=20$.}
\end{figure}

Molecular Dynamics simulations were performed by cooling the
liquid at constant pressure using Berendsen thermostat and barostat
during the equilibration phase. Standard Velocity-Verlet algorithm was
used to integrate the equations of motion. In order to achieve better
control on temperature in the deeply supercooled regime, we performed
a few production runs using the Nos\'e-Poincar\'e
thermostat~\cite{bond,nose01}. The timestep $\delta t$ was varied
between 0.002 at high temperature and 0.006 at low 
temperature. 
The time constant for the Berendsen
thermostat~\cite{book:at} was $t_T=\delta t/0.1$, while the coupling
constant the for Berendsen barostat~\cite{book:at} was $10^3$ in reduced units. 
The inertia parameter of the Nos\'e-Poincar\'e thermostat was set to
$Q=5$. Constant pressure simulations provide a means to 
compare different mixtures in a way similar to the one employed in
experiments. There is also interest in
understanding how the sampling of the energy landscape changes
when isobaric quenches are 
considered instead of isochoric ones~\cite{middleton03}. The density
variations along isobaric quenches at a pressure $P=10$ are shown for a
selection of mixtures in Figs.~\ref{fig:presrho1} and~\ref{fig:presrho2}. 
In order to make a comparison with standard constant density simulations, we
also performed some isochoric quenches for BMLJ and WAHN by
fixing the density at the value used in the original papers. A summary
of our thermal histories is shown in Table~\ref{tab:sim}. For further
details about quenching protocols see Ref.~\onlinecite{coslovich07a}. 

Description of the minimization technique employed for locating
stationary points of the PES requires some further
comments. We followed a simple and popular
approach~\cite{angelani03}, which consists of minimization of the mean
square total force $W$ of the system, using the L-BFGS
algorithm~\cite{nocedal}. For each state point, some hundreds of
independent configurations (typically between 200 and 1000) from
simulation runs were considered as starting points for
$W$-minimizations. Some care has to be taken, since this
numerical procedure often leads to quasisaddles, i.e., points with
small but non-zero $W$, which display an inflection
mode~\cite{doye02,wales03,angelani03}. As a criterion for
distinguishing true saddles we use $W\alt W_0 \equiv 10^{-12}$,
similarly to previous studies~\cite{angelani03,grigera06}.
The fraction of true saddles sampled in
small-sized samples ($N=108$) is rather large (from 10\% to 30\% for
WAHN, from 5\% to 20\% for BMLJ, 
depending on temperature), so that this approach appears to be quite
feasible for similar system sizes. For larger samples, the fraction of
true saddles decreases. However, from on overall point of view, our
findings indicate that saddles and quasisaddles share similar average 
properties, in agreement with several previous
observations~\cite{angelani03,sampoli03,shah03}. We also found that
ensuring continuity of the interaction potential up to the first
derivative at $r=r_c$, i.e., employing the QS cutoff, is enough to
avoid major round-off errors in $W$-minimizations. Moreover, the
fraction of true saddles sampled for $N=108$ does not increase
appreciably when using the smoother CSPL cutoff. 
 
\section{Effective activation energies}\label{sec:fragility}

A common way to display the temperature dependence of structural
relaxation times $\tau(T)$ in supercooled liquids is by
construction of the so-called Angell plot, in which $\log\tau$ is 
shown against $T_g/T$. 
In a previous work~\cite{coslovich07a}, we used a similar approach to analyze the
variations of fragility in Lennard-Jones mixtures. Here, we 
take a different view of the same problem
and analyze the effective activation (free) energy $E(T)$ defined by
inversion of  
\beq\label{eqn:actenedef} 
\tau(T)=\tau_{\infty}\exp\left[\frac{E(T)}{T}\right]
\eeq
where $\tau_{\infty}$ is the high-temperature limit of relaxation
times. Analysis of the temperature dependence of $E(T)$ will allow us
to make contact with previous work based on effective
activation energies~\cite{kivelson96}, and to discuss the variation of
fragility in a way more convenient for highlighting the role of the PES
(Sec.~\ref{sec:pes}).

Experimental and numerical analysis of effective activation energies
$E(T)$ in supercooled liquids clearly signals a crossover between two distinct
dynamical regimes. At high temperature, relaxation times 
follow a mild, Arrhenius-type temperature dependence, $\tau_{\infty}
\exp\left[E_{\infty}/T\right]$. Hence, in the normal liquid regime, we
have $E(T)\approx E_\infty$. Below some crossover temperature $T^*$,
effective activation energies of fragile glass-formers increase
markedly, indicating super-Arrhenius behavior. The more fragile is
the glass-former, the shaper the increase of effective activation
energies below $T^*$. 
The features of $E(T)$ just discussed are
incorporated in the so-called frustration-limited domains theory of the
glass-transition~\cite{tarjus04}, which predicts $E(T)$ to be of the form
\beq\label{eqn:actene} 
E(T) = 
\left\{ 
\begin{array}{ll}
  E_{\infty} & T>T^*   \\
  E_{\infty} + B T^* {\left( 1-\frac{T}{T^*} \right)}^{8/3} & T<T^* \\
\end{array}
\right.
\eeq
It has been shown~\cite{kivelson96} that the functional form in
Eq.~\eqref{eqn:actene} provides a fair account of a wide spectrum of
experimental data. Fragility is measured by $B$, which is the
parameter quantifying the departure from the high-temperature
Arrhenius regime. Further discussions on the role of the other
parameters in Eq.~\eqref{eqn:actene} and on the exponent 8/3 can be found in
Refs.~\onlinecite{ferrer99}~and~\onlinecite{tarjus00}. 

\begin{table*}
\caption{\label{tab:fragility}
  Parameters of fits to Eq.~\eqref{eqn:actene} for effective
  activation energies $E(T)$ of large and small species. The reference
  temperature $T_r$ and the onset temperature $T_{onset}$ are
  described in the text.}  
\begin{ruledtabular}
\begin{tabular}{lllllllllll}
    &    &                & \multicolumn{4}{c}{Large particles}
                          & \multicolumn{4}{c}{Small particles} \\
\cline{4-7}
\cline{8-11}
    & P   & $T_r$ & $T^{*}$  & $B$ & $\tau_{\infty}$ & $E_{\infty}$
                    & $T^{*}$  & $B$ & $\tau_{\infty}$ & $E_{\infty}$ \\
\hline
      BMLJ &       5 & 0.464 &    0.83(2) & 33 $\pm$ 5 &  0.0931(3) &    1.99(1) &    0.81(1) & 40 $\pm$ 5 &   0.097(1) &    1.62(3) \\ 
      BMLJ &      10 & 0.574 &    1.05(1) & 29 $\pm$ 3 &  0.0815(5) &    2.61(1) &    1.06(1) & 30 $\pm$ 2 &  0.0889(9) &    1.93(2) \\ 
      BMLJ &      20 & 0.765 &    1.41(4) & 26 $\pm$ 4 &   0.067(1) &    3.71(9) &    1.44(4) & 26 $\pm$ 4 &  0.0733(8) &    2.69(3) \\ 
      BMLJ &      50 & 1.248 &    2.27(5) & 28 $\pm$ 4 &  0.0481(9) &    6.60(7) &    2.35(5) & 27 $\pm$ 3 &   0.052(1) &     4.7(1) \\ 
      WAHN &      10 & 0.623 &    0.94(1) & 77 $\pm$ 11 &  0.0825(4) &    2.33(1) &    0.91(1) & 91 $\pm$ 12 &  0.0567(3) &    2.44(1) \\ 
      WAHN &      20 & 0.825 &    1.23(2) & 82 $\pm$ 15 &  0.0670(6) &    3.38(3) &    1.20(2) & 96 $\pm$ 15 &  0.0455(3) &    3.57(3) \\ 
 AMLJ-0.60 &      10 & 0.451 &    0.92(2) & 20 $\pm$ 2 &   0.076(1) &    2.43(3) &    0.90(1) & 18 $\pm$ 1 &  0.0828(7) &    1.69(2) \\ 
 AMLJ-0.64 &      10 & 0.474 &    0.88(1) & 27 $\pm$ 3 & 0.07691(3) &   2.444(1) &    0.88(1) & 25 $\pm$ 2 &  0.0834(4) &    1.76(1) \\ 
 AMLJ-0.70 &      10 & 0.514 &    0.86(2) & 46 $\pm$ 9 &  0.0811(1) &   2.359(4) &    0.84(2) & 49 $\pm$ 9 &   0.084(1) &    1.84(5) \\ 
 AMLJ-0.73 &      10 & 0.560 &    0.82(1) & 94 $\pm$ 16 &  0.0785(7) &    2.48(2) &    0.84(1) & 86 $\pm$ 14 &  0.0839(7) &    1.93(2) \\ 
 AMLJ-0.76 &      10 & 0.601 &    0.84(1) & 128 $\pm$ 22 &  0.0790(9) &    2.49(3) &    0.85(1) & 120 $\pm$ 20 &   0.083(1) &    2.00(5) \\ 
 AMLJ-0.82 &      10 & 0.636 &    0.90(1) & 100 $\pm$ 18 &  0.0803(5) &    2.53(1) &    0.92(1) & 92 $\pm$ 16 &  0.0829(8) &    2.17(2) \\ 
\hline
    & $\rho$ & $T_r$ & $T^{*}$  & $B$ & $\tau_{\infty}$ & $E_{\infty}$ 
                    & $T^{*}$  & $B$ & $\tau_{\infty}$ & $E_{\infty}$ \\
\hline
      BMLJ &     1.2 & 0.422 &    0.77(1) & 22 $\pm$ 3 &   0.110(2) &    2.69(2) &    0.85(1) & 17 $\pm$ 2 &  0.0909(7) &    2.25(1) \\ 
      WAHN &     1.3 & 0.522 &    0.81(1) & 65 $\pm$ 6 &   0.097(1) &    2.73(2) &    0.81(1) & 61 $\pm$ 6 &   0.071(1) &    2.70(3) \\ 
\end{tabular}

\end{ruledtabular}
\end{table*}

The dynamical quantity on which we focus in this section is the
relaxation time for the decay of density fluctuations, as identified by the self
part of the intermediate scattering function $F_s^{\alpha}(k,t)$, where
$\alpha=1,2$ is an index of species. We define relaxation times
$\tau_{\alpha}$ by requiring that $F_s^{\alpha}(k^*,t)$
has decayed to $1/e$, where $k^*\approx 8$ 
is the position of the first maximum in the number-number static
structure factor of the mixtures in consideration~\cite{coslovich07a}.
A first guess of the crossover temperature $T^*$ is provided by the
temperature $T_{onset}$ at 
which two-step, non-exponential relaxation of $F^{\alpha}_s(k,t)$ is
first observed upon cooling the liquid from high 
temperature~\cite{ka2}. For fitting our data to Eq.~\eqref{eqn:actene}, we 
proceed as suggested by Kivelson~\etal~\cite{kivelson96}. First we
fit the high temperature portion of our data ($T>T_{onset}$) to an
Arrhenius law $\tau_{\infty} 
\exp\left[E_{\infty}/T\right]$ and then we use $\tau_{\infty}$ and
$E_{\infty}$ as fixed parameters in global a fit of our simulation
data to Eq.~\eqref{eqn:actene}. Note that, although $T^*$ is
considered as a fitting parameter, its final value is never too far
from the initial guess $T_{onset}$. 

\begin{figure}
\includegraphics*[width=0.46\textwidth]{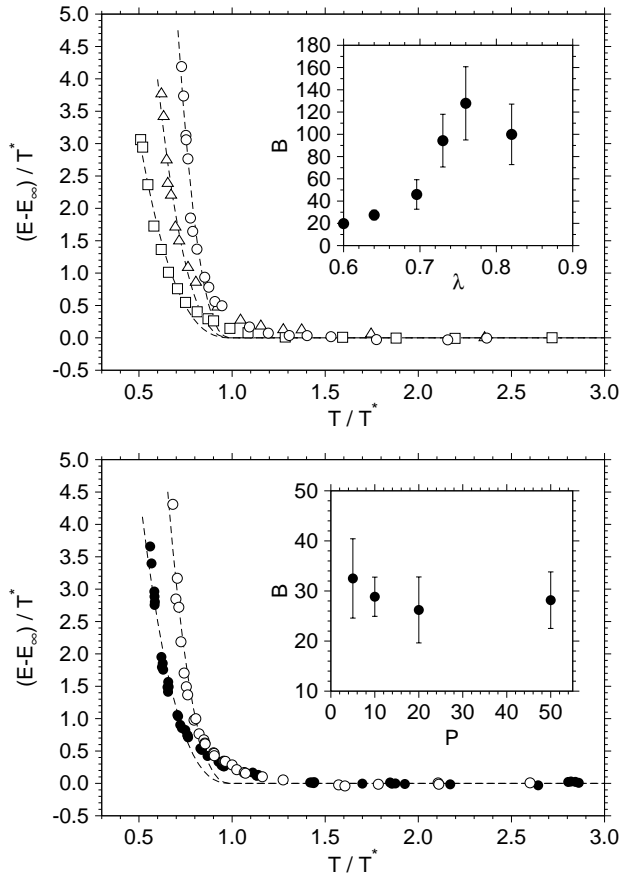}
\caption{\label{fig:barrtau1}
  Effective activation energies for relaxation of large particles,
  after removal of the high-temperature limit $E_{\infty}$. The
  dependence of $(E(T)-E_{\infty})/T^*$ on reduced temperature $T/T^*$
  is shown along isobaric quenches at $P=10$. 
  Dashed lines are fits to Eq.~\eqref{eqn:actene}. Upper plot: AMLJ-$\lambda$
  mixtures for values of size ratio $\lambda=0.60$ (squares) ,$0.70$
  (triangles) and $0.76$ (circles). Lower plot: BMLJ (filled circles)
  and WAHN (open symbols). Inset of upper plot: fragility index $B$ of
  AMLJ-$\lambda$ versus $\lambda$. Inset of lower plot:
  fragility index $B$ versus $P$ obtained for different
  isobars in BMLJ.}       
\end{figure}

In Table~\ref{tab:fragility}, we summarize the results of our fitting
procedure for Eq.~\eqref{eqn:actene}. Considering separately the cases
of effective activation energies for large and small particles, we
find that the fitted parameters for the two species show similar
trends of variation in different systems and for different pressure
and density conditions. In the following, we will thus simply focus on the
effective activation energies $E(T)$ obtained from the relaxation
times $\tau\equiv\tau_1$ of large particles.   
In order to put into evidence the variation of fragility index $B$ for
different mixtures we plot, as in Ref.~\onlinecite{tarjus04},
the difference $(E(T)-E_{\infty})/T^*$ against the reduced temperature 
$T/T^*$. In Fig.~\ref{fig:barrtau1}, we show results obtained along
isobaric quenches at $P=10$ for a selection of AMLJ-$\lambda$ mixtures
(upper plot), and for BMLJ and WAHN mixtures (lower plot).  
The first important point is that there is a systematic variation of
fragility with size ratio. Below the crossover temperature $T^*$
effective activation energies increase faster as the size ratio
$\lambda$ increases, i.e., AMLJ-$\lambda$ mixtures become more fragile
as $\lambda$ increases. The second point is that the BMLJ mixture is less
fragile than the WAHN mixture. These observations are substantiated by the
outcome of our fitting procedure. From an
overall point of view, we find that Eq.~\eqref{eqn:actene} provides a
good fitting function for our simulation data. Actually, the crossover
around $T^*$ in our simulation data is smoother than predicted by
Eq.~\eqref{eqn:actene}, but it should also be remarked that
Eq.~\eqref{eqn:actene} is not expected to hold exactly around
$T^*$~\cite{kivelson96}. In the inset of the upper plot of
Fig.~\ref{fig:barrtau1}, the fragility parameter $B$ is shown as a
function of size ratio for AMLJ-$\lambda$ mixtures. Despite the
somewhat large uncertainty on our estimate of 
$B$, there is a clear trend of increase of $B$ as $\lambda$ increases and
a tendency to saturate around $\lambda\approx 0.80$. 
Results obtained along different isobars for BMLJ show that the isobaric
fragility index $B$ for this system is essentially pressure invariant in the range
$5\leq P \leq 50$, as it can be seen from the inset of the lower plot
in Fig.~\ref{fig:barrtau1}. 

\begin{figure}
\includegraphics*[width=0.46\textwidth]{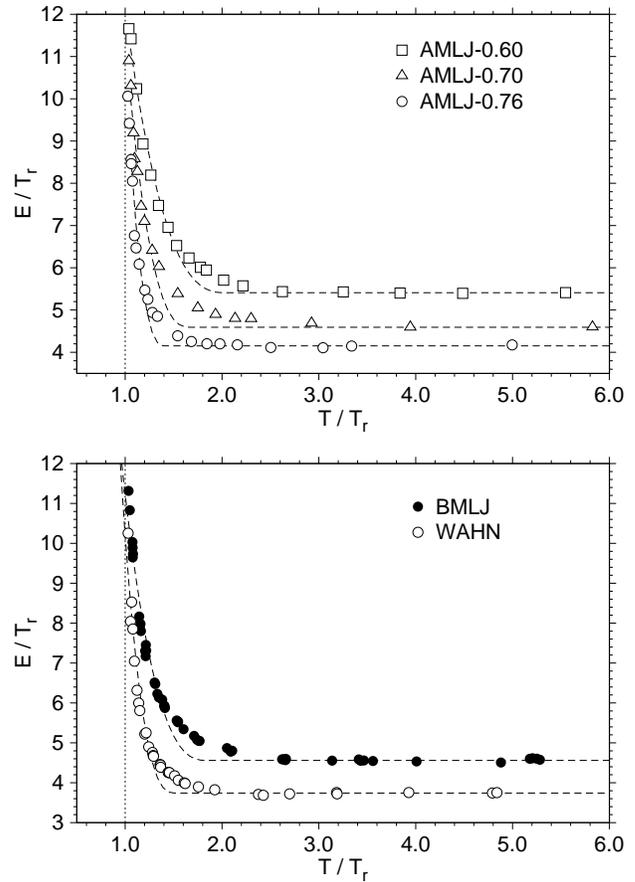}
\caption{\label{fig:barrtau3}
  Scaled effective activation energies $E/T_r$ as a function of
  $T/T_r$, along the isobar $P=10$. Dashed lines are fits to
  Eq.~\eqref{eqn:actene}. Upper plot: AMLJ-$\lambda$ mixtures, for
  $\lambda=0.60$ (squares), $\lambda=0.70$ (triangles) and
  $\lambda=0.76$ (circles). Lower plot: BMLJ (filled circles) and WAHN
  (open circles).}        
\end{figure}

Within the frustration-limited domains theory~\cite{tarjus04}, the
crossover temperature $T^*$ is interpreted as an intrinsic ridge between two distinct dynamical
regimes, and should thus provide a means to scale and compare properties of
different glass-formers.  
Nonetheless, the use of isochronic, conventional reference
temperatures, such as the glass-transition temperature $T_g$, is often
useful and effective~\cite{ferrer99}. We thus introduce a reference
temperature $T_r$ such that the relaxation time for large particles
reaches $\tau(T_r)=4\times 10^4$. 
Since the value $\tau_{\infty}$ obtained from the high-temperature
behavior is roughly system independent at a given pressure, the
activation energies for different 
systems converge to a common value $E(T_r)/T_r \approx
\log(\tau(T_r)/\tau_{\infty})\approx 12$ as $T\rightarrow T_r$. Thus,
a plot in which both $E(T)$ and $T$ are scaled by $T_r$ can be
regarded as a generalized Angell plot, in which activation
energies for different systems, when considered along the \bem same
\eem isobar, 
converge around $T_r$ to a common value. This kind of plot is shown in
Fig.~\ref{fig:barrtau3}, where we compare AMLJ-$\lambda$ mixtures for  
different values of $\lambda$ (upper plot) and the two prototypical
mixtures BMLJ and WAHN (lower plot). In this representation, fragility can be 
seen from a more pronounced increase of effective activation energies,
relative to the high temperature limit. A rough 
estimate of fragility can be thus obtained from the value of
$E_{\infty}/T_r$, a fact which resembles the experimental correlation
between $E_{\infty}/T_g$ and fragility~\cite{novikov05}. 

A comparative analysis, based on Eq.~\eqref{eqn:actene}, of
experimental and numerical data was attempted some years ago by
Ferrer~\etal~\cite{ferrer99}, and further discussed by  
Tarjus~\etal~\cite{tarjus00}. The outcome of the fitting procedure
led these authors to raise some doubts about the fragile nature of
some simulated models of supercooled liquids, including the BMLJ
mixture. This was contrary to the expectation, based on qualitative
grounds~\cite{angell99}, that Lennard-Jones mixtures should be fragile
glass-formers.  
Given the variety of Lennard-Jones models and external conditions analyzed
in this work, we are probably in the position to shed some light on
this point.  
First, we note that, for \bem all \eem mixtures considered, the ratio
$E(T)/E_{\infty}$ is already larger than 2 around $T_r$. We remark
that this is a typical fragile behavior, even when compared to
experimental data for fragile glass-formers such as
ortho-terphenyl~\cite{tarjus00}. Note that, a part from the   
trivial determination of $E_{\infty}$, this results is independent on
the fitting procedure. Second, comparisons between experiments and
numerical simulations of supercooled liquids should always be made with
care. A much more limited temperature range is available in numerical
simulations, and this can bias the results of fits to
Eq.~\eqref{eqn:actene}. For instance, by restricting the 
temperature range for fitting so that $\tau\alt 10^2$, we obtained for BMLJ
values of fragility index as low as $B\approx 12$ at constant pressure, and
$B\approx 4$ at constant density, in line with the results 
obtained in Ref.~\onlinecite{tarjus00} by considering a similar range of
$\tau$. Fitting our data down to $T_r$, we obtain $B\approx 30$ for
BMLJ at constant pressure, and we expect 
that equilibrating the system at even lower temperatures would yield
slightly larger values of $B$. Also note that for additive 
Lennard-Jones mixtures with moderate size asymmetry we find $B\approx
100$, which is already typical of intermediately fragile liquids
($B\approx 90$ for glycerol~\cite{tarjus00}).   
Thus, from an overall point of view, Lennard-Jones mixtures
appear to be fragile glass-formers, as may be expected for simple
systems with non-directional interactions. On the other hand, it is
true that some Lennard-Jones mixtures are less fragile than others. In
particular, the well-studied BMLJ mixture, is not among the most
fragile Lennard-Jones mixtures. 

\section{Potential energy surface}\label{sec:pes}

An important role for
understanding the dynamical features of supercooled liquids is played
by the stationary points of the PES and, more generally, by the negatively
curved regions of the
PES~\cite{angelani00,angelani02,angelani03,sampoli03}. 
In this section, we will adopt a simple, non-topographic point of view,
ignoring the connectivity of stationary points. Approaches based on pathways
connecting adjacent minima through transition states
\cite{middleton01b,middleton03}, or transitions between
metabasins \cite{doliwa03,doliwa03a,doliwa03b} have been 
recently developed and applied to some model supercooled liquids, but
they require expensive and complex numerical procedures. 
We will thus focus on some simple statistical features of the PES,
and discuss their correlations to fragility in Lennard-Jones mixtures. 

In the following, we will investigate the local curvature of the
PES by looking at the Hessian matrix $\mathcal H$ of the potential
energy. Standard diagonalization of $\mathcal H$ yields a set of $3N$ modes
with eigenvalues $\omega^2_\alpha$ and eigenvectors
$\eve_j^{\alpha}$, where $\alpha=1,\dots,3N$ is an index of mode and
$j=1,\dots,N$ is an index of particle. Modes are then 
classified as stable if $\omega^2_\alpha$ is positive (real frequency), or
unstable if $\omega^2_\alpha$ is negative (imaginary frequency). For liquids,
most of the relevant information is encoded in the 
unstable modes of the PES, whose analysis usually comes in two varieties  
\footnote{Actually, also a third way has been considered~\cite{chowdhary02}.}. 
The first approach is referred to as
Instantaneous Normal Modes (INM) analysis, and 
focuses on instantaneous configurations sampled along the MD trajectory
\cite{bembenek,gezelter,keyes97}. 
The second approach considers high-order 
stationary points of the PES, obtained using minimization
procedures~\cite{angelani00,broderix00,doye02,wales03,angelani03}. 
According to the number of unstable modes $n_u$ in the Hessian matrix,
stationary points are classified as local minima ($n_u=0$) 
or saddles ($n_u\neq 0$). As mentioned in Sec.\ref{sec:model},
quasisaddles are other points of the PES often
reached by the minimization algorithm employed in
this work. The exclusion of quasisaddles from statistical
averages will not affect the main conclusions of this section. We
checked the reliability of our results on some smaller samples of
$N=108$ particles, in which a larger fraction of saddles could be found
(Sec.~\ref{sec:model}). In the following, we will focus on the larger
sample ($N=500$) and we will mostly use the term saddles in a broad
sense, without distinction between saddles and quasisaddles.  

\begin{figure}
\includegraphics*[width=0.46\textwidth]{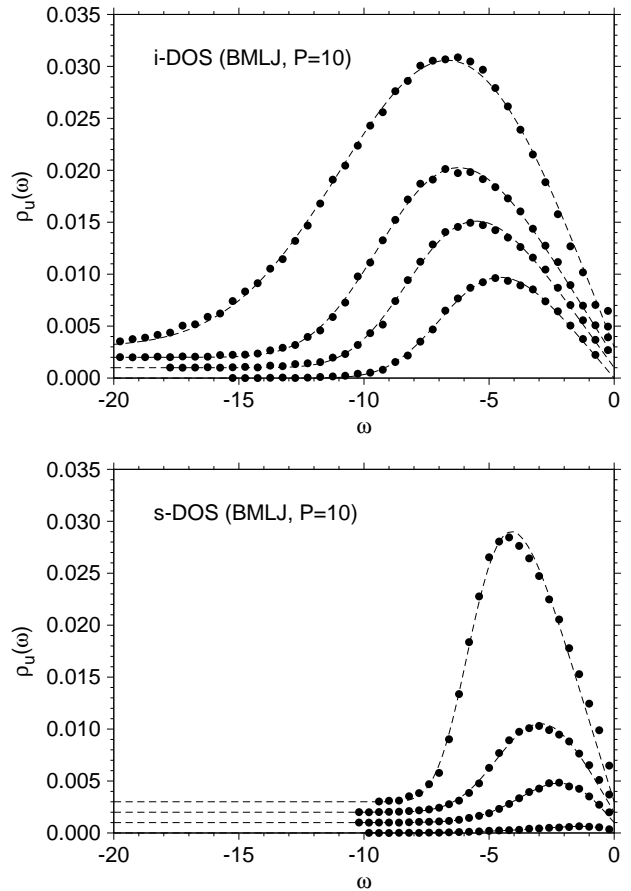}
\caption{\label{fig:vdos}
  Unstable branch of density of states $\rho_u(\omega)$ for
  instantaneous configurations (i-DOS, upper plot) and saddles (s-DOS,
  lower plot) in BMLJ. Results are shown at four different state
  points  at $P=10$ (from top
  to bottom, $T=2.0,1.0,0.7,0.6$). Dashed lines are fits to
  Eq.~\eqref{eqn:vdos}.} 
\end{figure}

As a starting point, we consider the ensemble-averaged density of
states (DOS)
\beq\label{eqn:vdosdef}
\rho(\omega;T) = \Big\langle \sum_{\alpha=1}^{3N} 
                           \delta(\omega_\alpha-\omega) \Big\rangle_T
\eeq
at temperature $T$. The thermal average in Eq.~\eqref{eqn:vdosdef} can be
performed using either instantaneous configurations (i-DOS) or saddles
(s-DOS). The unstable branch of $\rho(\omega;T)$ will be denoted by
$\rho_u(\omega;T)$, and imaginary frequencies will be shown, as usual,
along the real negative axis. To provide a quantitative account of
the features of $\rho_u(\omega;T)$, we consider the following
functional form 
\beq \label{eqn:vdos}
\rho_u (\omega; T) = A\,\omega\, \exp\!\left[{\left(\frac{B\omega}{\sqrt{T}}\right)}^C\right]
\eeq
which has been shown to describe well the unstable
i-DOS~\cite{keyes97}. Specific functional forms for the s-DOS have been discussed, for 
instance, in the context of $p$-spin models~\cite{cavagna01b,cavagna03}, but
they fail to meet some basic requirements for realistic
systems, such as the behavior $\rho(\omega)\rightarrow 0$ for
$\omega\rightarrow 0$. We have thus attempted to apply Eq.~\eqref{eqn:vdos} also to
s-DOS and found that Eq.~\eqref{eqn:vdos} provides an excellent fit
for both i-DOS and s-DOS, becoming inadequate only at very high 
temperature or in the limit of vanishing interval of imaginary
frequencies. The quality of the fits obtained using
Eq.~\eqref{eqn:vdos} is exemplified in Fig.~\ref{fig:vdos} for
different state points of BMLJ.  

\begin{figure}
\includegraphics*[width=0.46\textwidth]{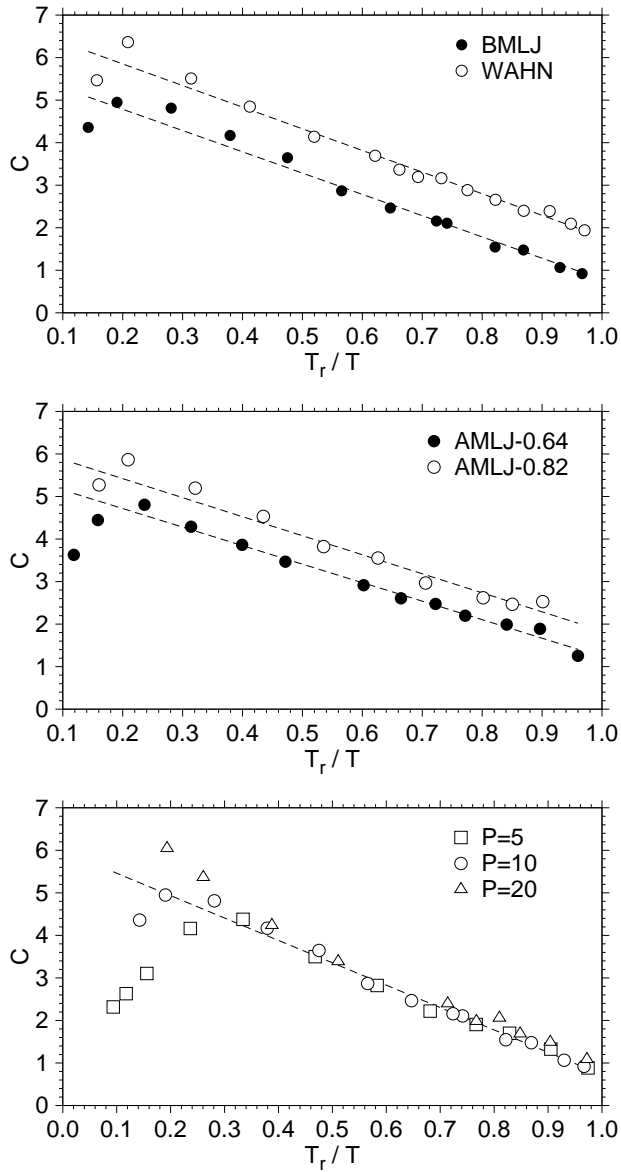}
\caption{\label{fig:vdosC}
  Parameter $C$ for s-DOS obtained from fits to Eq.~\eqref{eqn:vdos} as a
  function of $T_r/T$. Upper plot: BMLJ (filled circles) and WAHN
  (open circles) at $P=10$, Middle plot: AMLJ-0.64 (filled circles) and
  AMLJ-0.82 (open circles) at $P=10$. Lower plot: BMLJ at different
  pressures, $P=5$ (squares), $P=10$ (circles) and $P=20$
  (triangles). Dashed lines are fits of the type $a+b(T_r/T)$, with
  $b\approx 5$ being roughly system independent. 
.} 
\end{figure}

In the context of the INM theory of
diffusion~\cite{keyes95a}, super-Arrhenius behavior is 
explained in terms of the temperature dependence of the parameters in
Eq.~\eqref{eqn:vdos} for the i-DOS, and should be primarily signaled by an
increase of $C$ by decreasing temperature
\cite{nitzan95,keyes97,bembenek,li99}. 
Unfortunately, we found that INM theory is not able to put into evidence
the different fragility of the Lennard-Jones mixtures considered in this
work~\footnote{Introduction of a lower frequency cutoff $\omega_c$ to
filter some shoulder modes~\cite{li99} would not affect our
conclusions.}. For instance, we found that $C$, as a function of
$T/T_r$, has similar values in all systems, whereas we would have
expected a sharper increase in the case of more
fragile mixtures. Analysis of the s-DOS in terms of Eq.~\eqref{eqn:vdos}
provides a different, sharper picture. Without attempting a detailed
analysis of the temperature dependence of all parameters
in Eq.~\eqref{eqn:vdos}, we will focus on the behavior of parameter
$C$. In Fig.~\ref{fig:vdosC}, we show the dependence of $C$ on $T/T_r$ for
the s-DOS of different mixtures at constant pressure. Apart from some
deviations at very high-temperature, $C$ decreases by decreasing
temperature, differently from
the case of the i-DOS. In all systems, we observe a temperature dependence of the
type $C \sim b (T_r/T)$. Interestingly, all mixtures seem to share a
common value of the slope $b\approx 5$ in a plot of $C$ versus $T_r/T$,
and we find a shift towards larger values of $C$ as
fragility increases. As a check of the relation between $C$ and
fragility, the values of $C$ along different isobars for BMLJ collapse on a master
curve when plotted against $T/T_r$, consistently with the pressure
invariance of fragility reported in Sec.~\ref{sec:fragility}. Our
results thus indicate that Eq.~\eqref{eqn:vdos} could 
provide a good starting point for modeling 
the s-DOS in realistic systems and that a saddle-based approach is
more sensitive to the dynamical behavior of supercooled Lennard-Jones
mixtures than INM. 

\begin{figure}[!ht]
\includegraphics*[width=0.44\textwidth]{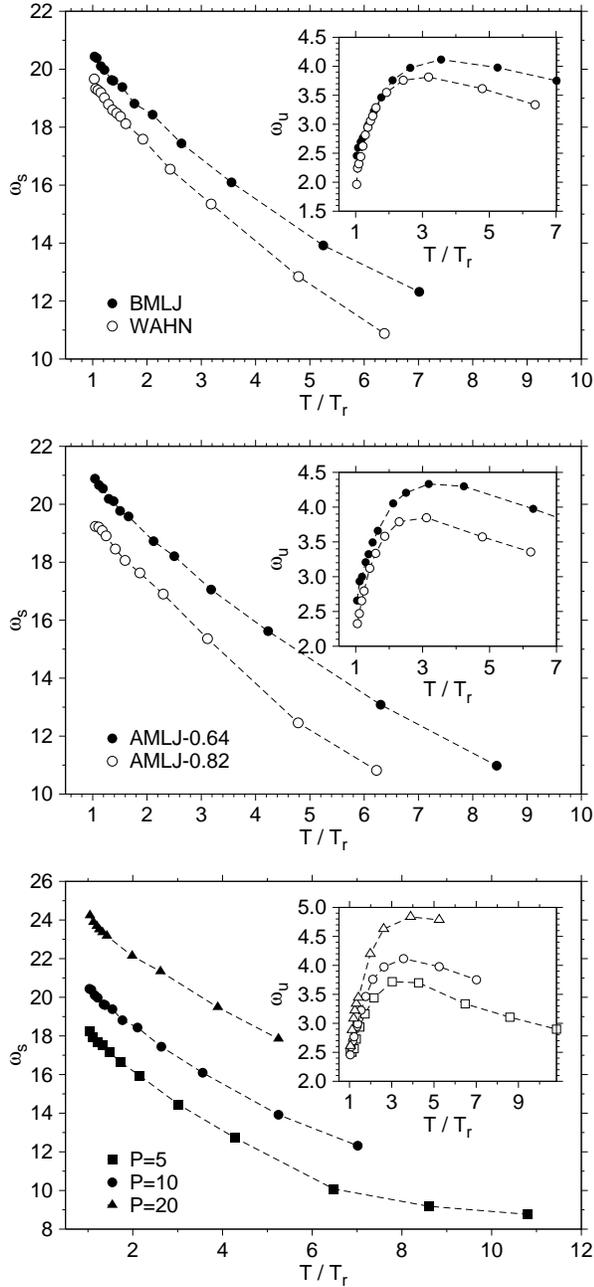}
\caption{ \label{fig:sadw}
  Average frequency of stable modes $\omega_s$ (main plots) and
  unstable modes $\omega_u$ (insets) of saddles as a function of 
  $T/T_r$.  
  Upper plot: BMLJ (filled circles) and
  WAHN (open circles) at $P=10$. 
  Middle plot: AMLJ-0.64 (filled circles) and
  AMLJ-0.82 (open circles) at $P=10$. 
  Lower plot: BMLJ at $P=5$ (squares),
  $P=10$ (circles), and $P=20$ (triangles). 
  } 
\end{figure}

In the light of previous studies of vibrational properties of local
minima~\cite{sastry01,middleton01b}, it might be asked whether
coarse-grained quantities obtained from the s-DOS,
such as the average frequency of of stable modes 
\beq
\omega_s(T) = \int_{-\infty}^0 d\omega\,\omega \rho(\omega;T)
\eeq
and unstable modes 
\beq
\omega_u(T) = \int_0^{\infty} d\omega\,\omega \rho(\omega;T)
\eeq
already convey information about fragility.
In Fig.~\ref{fig:sadw}, we show the dependence of
$\omega_u$ and $\omega_s$ on $T/T_r$ for different mixtures at constant
pressure. We found analogous thermal behaviors by considering
geometric mean frequencies of stable and unstable modes. Similarly to
what happens in the case of local minima~\cite{middleton03},
constant-pressure data show an increase of $\omega_s$ by decreasing
temperature, i.e., by decreasing energy of saddles. This 
behavior is opposite to the one observed in constant-density
simulations. The average frequency of unstable modes $\omega_u$
always shows a non-monotonic temperature dependence, 
characterized by a maximum at intermediate temperatures, 
which is peculiar to isobaric quenches. 
Comparing mixtures along isobaric quenches at $P=10$, we find a slight
shift to larger absolute vibrational frequencies, as fragility
decreases.   
However, the robustness of this correlation is weakened when it is
tested using the pressure invariance of isobaric fragility in BMLJ.  
In the bottom plot of Fig.~\ref{fig:sadw}, we look at the behavior of
$\omega_s$ and $\omega_u$ along different isobars in BMLJ. As the
pressure $P$ of the isobar increases, vibrational frequencies are shifted markedly
to larger absolute values, most probably by the increasing
density~\footnote{It has been suggested~\cite{middleton03} that the 
leading contribution to the density dependence of the geometric mean frequency in
local minima should scale as a power law of
$\rho$. We argue that a similar argument might
hold for high-order stationary points.}. This behavior led us to reconsider the case of
local minima along different isobars in BMLJ, and we found a similar trend in
vibrational properties. Thus, although some correlation 
might be observed at a given pressure, there seems to be no direct
connection between average vibrational frequencies of stationary
points and fragility. 

\begin{figure}[!ht]
\includegraphics*[width=0.46\textwidth]{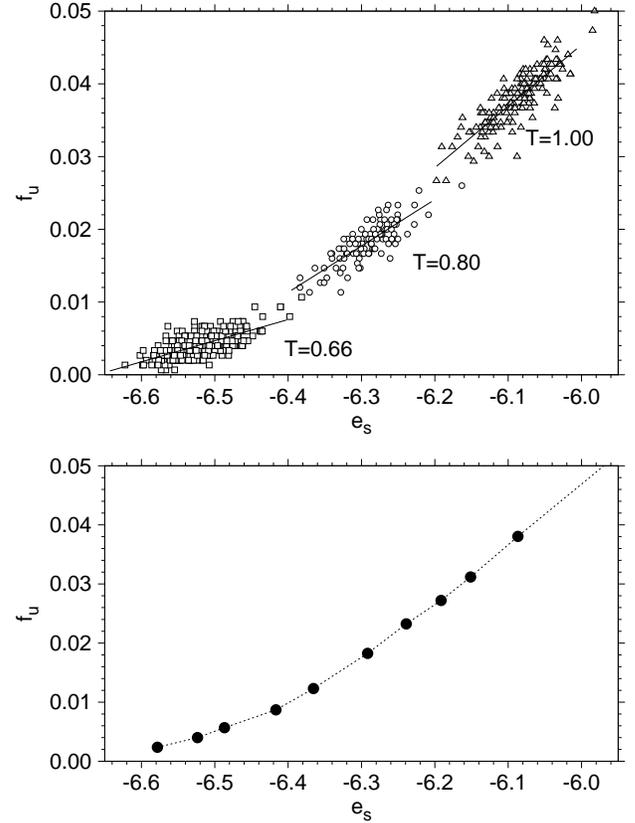}
\caption{\label{fig:sad1}
  Upper plot: scatter plot of the fraction of unstable modes
  against energy of single saddles. Results are shown for WAHN at
  $P=10$ for three different state points: $T=0.66$ (squares),
  $T=0.80$ (circles) and $T=1.00$ (triangles). Linear
  fits of the type $f_u=a+b e_s$ (solid lines) are used to estimate
  the derivative in Eq.~\eqref{eqn:sadbar}, i.e., $E_s=1/3b$. Lower
  plot: parametric plot of average unstable modes of saddles $f_u(T)$
  against energy of saddles $e_s(T)$, for WAHN at $P=10$. 
} 
\end{figure}

\begin{figure*}[ht]
\includegraphics*[width=0.94\textwidth]{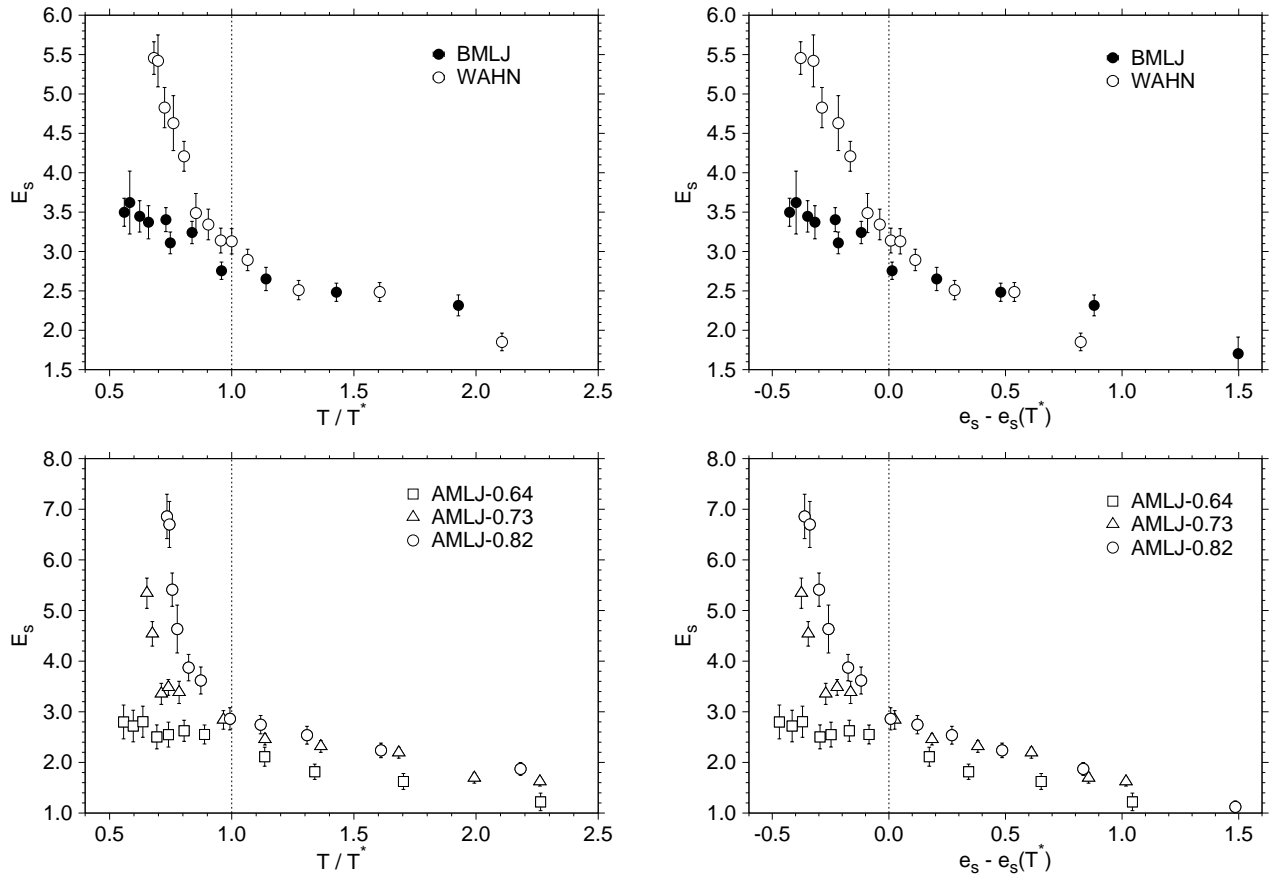}
\caption{\label{fig:sadbar}
  Effective energy barriers $E_s$ as a function of reduced temperature
  $T/T^*$ (left column) and and as a function of $e_s-e_s(T^*)$ (right
  column). Upper plots: WAHN (open circles) and BMLJ
  (filled circles) at $P=10$. Lower plots: AMLJ-0.82 (open circles)
  and AMLJ-0.64 (filled circles) at $P=10$.}  
\end{figure*}

The variation of fragility in Lennard-Jones mixtures, as discussed in
terms of effective activation energies for relaxation $E(T)$,
calls for an explanation based on energy barriers. Whereas it is
clear that $E(T)$ in Eq.~\eqref{eqn:actenedef} is rather an
activation \bem free \eem energy, the leading contribution to it might
already come from potential energy barriers between single
saddles. 
To address this point, we follow the simple proposal of
Cavagna~\cite{cavagna01c}. The starting point is the relation 
\beq\label{eqn:fies}
f_u=f_u(e_s)
\eeq
between the fraction of unstable modes and the energy of
saddles. Eq.~\eqref{eqn:fies} will be treated as parametric in 
$T$, i.e., we consider the average fraction of unstable modes
\beq\label{eqn:fu}
f_u = f_u(T) = {\langle n_u/3N \rangle}_T 
             = \int_{-\infty}^0 d\omega \rho_u(\omega;T)
\eeq
and the average energy of saddles
\beq\label{eqn:es}
e_s = e_s(T)
\eeq
at temperature $T$. It has been shown that Eq.~\eqref{eqn:fies}, as
obtained from numerical simulations, is insensitive to the actual
minimization algorithm employed~\cite{grigera06}, and to the
inclusion of quasisaddles~\cite{sampoli03}. According to
Cavagna~\cite{cavagna01c}, the average energy difference   
\beq\label{eqn:sadbar}
E_s(e_s)=\frac{1}{3}\der{e_s}{f_{u}}
\eeq
between saddles of order $n$ and $n+1$ provides an estimate of
potential energy barriers in the PES. More refined treatments
would take into account the connectivity of saddles and existence of
a distribution of energy barriers~\cite{doye02}. 
In order to find $E_s(T)$, we compute the derivative in Eq.~\eqref{eqn:sadbar} 
by linear regression of $e_s$ vs. $f_u$ scatter data of saddles
sampled at temperature $T$, as illustrated in Fig.~\ref{fig:sad1}.

The temperature dependence of the effective energy barriers $E_s(T)$ is
shown in the left plots of Fig.~\ref{fig:sadbar} for different mixtures at constant
pressure. Below $T^*$, i.e., in the range of temperature where
activated dynamics is expected to become important~\cite{doliwa03a,berthier03}, the 
behavior of $E_s(T)$ correlates to the 
fragility of the mixture. In fact, the increase of effective energy
barriers upon supercooling is sharper and more pronounced, the more
fragile is mixture. In the case of the more fragile mixtures, we find
a striking similarity between the increase of $E_s(T)$ below $T^*$ and
that of the effective activation energies $E(T)$ defined by
Eq.~\eqref{eqn:actenedef}. In WAHN, for instance, we find
$E(T_r)\approx E_s(T_r)\approx 12 T_r$.  
The trends just discussed are in line with the results obtained by
direct calculations of energy barriers between adjacent 
minima in the soft-sphere version of WAHN~\cite{grigera02} and
BMLJ~\cite{doliwa03a}. 
Some concerns might regard the fact that $e_s(T)$, i.e., the energies
of saddles sampled at a given $T$, can depend on the 
minimization algorithm~\cite{grigera06}. On the other hand, the results 
obtained in Ref.~\onlinecite{grigera06} indicate that 
the \bem energy \eem dependence of the properties of saddles is much
less sensitive to the details of the minimization procedure employed.
We have thus analyzed our data for $E_s$ as a function of $e_s$, where
$e_s$ is given by the thermal average in Eq.~\eqref{eqn:es}, focusing
on the energy range below $e_s(T^*)$. For convenience, we 
have shifted the energies $e_s$ by $e_s(T^*)$. Such a representation
of our data is shown in the right plots of Fig.~\ref{fig:sadbar} and
confirms the trends discussed above on the basis of the temperature
dependence of $E_s$. Thus, independent of the representation used,
the average energy barriers show a strong connection to the variations
of fragility in our models. This also provides evidence of
the relevance, for the supercooled dynamics, of the $e_s(T)$ mapping
obtained through $W$-minimizations. 

\begin{figure}[!tb]
\includegraphics*[width=0.45\textwidth]{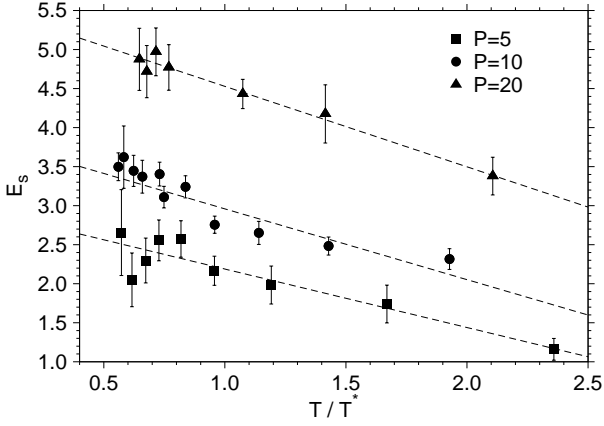}
\caption{\label{fig:sadbarP}
  Effective energy barriers $E_s$ as a function of reduced temperature
  $T/T^*$ for BMLJ at $P=5$ (squares), $P=10$ (circles), and $P=20$
  (triangles). Dashed lines represent linear fits.}  
\end{figure}

What is the effect of pressure on energy barriers? 
From the plot in Fig.~\ref{fig:sadbarP}, we see that
increasing pressure in BMLJ leads to larger potential energy 
barriers. This behavior is consistent with the results obtained by
Middleton and Wales~\cite{middleton03}, who calculated the 
distribution of potential energy barriers for diffusive rearrangements 
at different pressures for BMLJ.
What is made clear by our results, is that, at least in the case of
BMLJ, the increase of potential energy barriers with pressure has
little dynamical impact, because it is compensated by the increase of
the reference temperature $T_r$. That is, larger energy barriers will
be sampled at higher temperatures. Starting from data along
different isobars in BMLJ, in fact, we could obtain a rough master curve by
scaling both $E_s$ and $T$ by $T_r$. 

\begin{figure}[!tb]
\includegraphics*[width=0.44\textwidth]{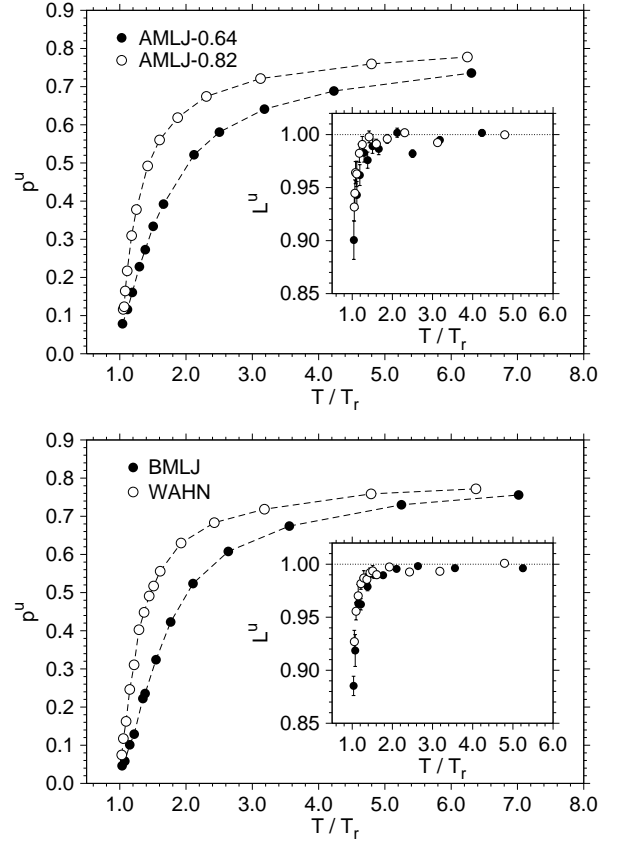}
\caption{\label{fig:sadloc1}
  Participation ratio $p^u$ of average squared displacements on
  unstable modes as a function of reduced
  temperature $T/T_r$ at $P=10$. Insets show the reduced gyration
  radius $L_u$ against $T/T_r$. Upper plot:
  AMLJ-0.82 (white squares) and AMLJ-0.64 (black squares). Lower plot:
  WAHN (white circles) and BMLJ (black circles).}
\end{figure}

\begin{figure}[!tb]
\includegraphics*[width=0.44\textwidth]{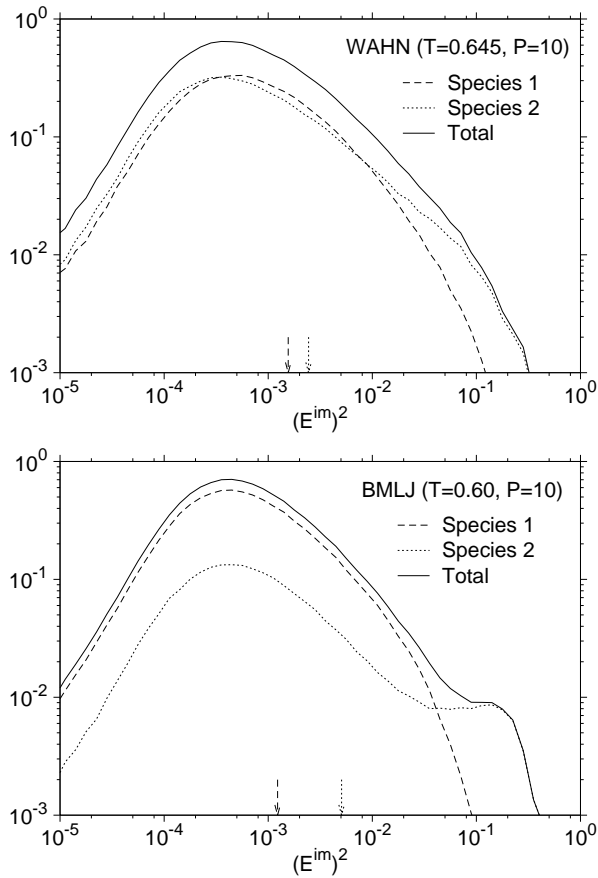}
\caption{\label{fig:sadmmsd1}
  Distribution of average squared displacements ${(E_i^u)}^2$
  on unstable modes of saddles. Results are shown for small particles
  (dotted lines), large particles (dashed lines), and irrespectively of chemical
  species (solid lines). Normalization is such that the area under each 
  curve is proportional to the corresponding number concentration. Arrows indicate
  the average values of ${(E_i^u)}^2$ for large and small
  particles. Upper plot: WAHN at $T=0.645$, $P=10$. Lower plot: BMLJ
  at $T=0.60$, $P=10$.} 
\end{figure}

The unstable modes of saddles sampled in the supercooled regime may
provide information about the elementary dynamical 
processes in the system~\cite{coslovich06}. It is thus of interest to
analyze the spatial localization features of unstable modes in
different Lennard-Jones mixtures, and see how they relate to
fragility.   
To address this point, we first average the squared
displacements for each particle over all the unstable modes~\cite{coslovich06}
\beq
{E_i^{u}}^2 = \frac{1}{n_{u}}\sum_{\alpha=1}^{n_{u}} {\eve_i^{\alpha}}^2
\eeq
Two different measures of localization for the vector of
average squared displacements ${E_i^{u}}^2$ are then considered.
The reduced gyration radius is defined as
\beq\label{eqn:lc}
L^u = \frac{1}{L/2} \left[ \sum_{i=1}^{N} |\rve_i - \rve_g|^2 {E_i^{u}}^2
  \right]^{1/2}
\eeq
where $L$ is the side of the simulation box. This quantity equals 1
when the vector ${E_i^{u}}^2$ is extended over the whole system, and
decreases progressively as the spatial localization of ${E_i^{u}}^2$
becomes more pronounced. The participation ratio is defined as
\beq\label{eqn:pu}
p^{u} = \left( N \sum_{i=1}^N {E_i^{u}}^4 \right)^{\!\!-1}
\eeq
and provides a rough estimate of the fraction of particles having
significant displacements in ${E_i^{u}}^2$. For instance, $p^u$ should
be $O(1)$ when the unstable modes are homogeneously distributed in the
system. 
The temperature dependence of these two quantities is shown in
Fig.~\ref{fig:sadloc1} for different Lennard-Jones mixtures. 
The existence of a sharp localization of unstable modes around $T_r$,
as identified by the abrupt decrease of $L^u(T)$, appears to be
a universal feature of saddles sampled by supercooled Lennard-Jones
mixtures. On the other hand, the pattern of localization of
the unstable modes changes according to the fragility of the
mixture. The more fragile is the mixture, the 
more rapid the localization of unstable modes upon supercooling, as
it is suggested by the behavior of $p^u(T)$. In the
range of temperature above $T_r$, we find that fragile mixtures tend to
have larger values of $p^u$. In this case, a larger fraction of
particles is thus involved in the unstable modes, which is consistent with
expectation that rearrangements should involve larger clusters as
fragility increases~\cite{jagla01}. We found further support to these
considerations by analyzing the average participation ratio and
gyration radius of individual unstable modes of saddles. 

\begin{figure*}[!ht]
\includegraphics*[width=0.91\textwidth]{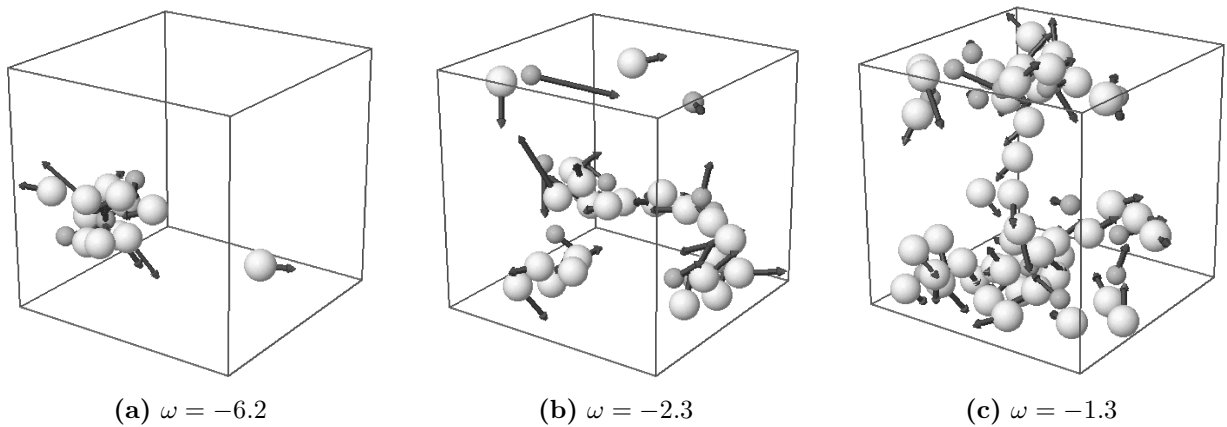}
\caption{\label{fig:modes2}
  Selection of three unstable modes of a quasisaddle ($n_u=4$) sampled
  in BMLJ at $T=0.66$, $P=10$. The nearly-zero mode of the
  quasisaddle has been ignored. A fourth unstable mode, not shown, is
  very similar in extension and shape to that shown in (b). For clarity,
  only particles having square displacements $(\eve_i^{\nu})^2$ larger than
  0.004 are shown, and eigenvectors are scaled
  logarithmically. Large and small particles are 
  shown as pale large spheres and small darker spheres,
  respectively. Note the strong localization of mode (a) and the
  existence of distinct string-like instabilities of large
  particles in mode (b) and (c).}   
\end{figure*}

Inspection of animated unstable modes of saddles sampled at low
temperature indicates the occurrence in BMLJ of strongly localized,
high-frequency unstable modes, in which few small particles
show very large displacements. This feature is reflected in a clearly bimodal
distribution of ${E_i^u}^2$ for small particles in deeply supercooled
BMLJ. 
In Fig.~\ref{fig:sadmmsd1}, we show the distributions of ${E_i^u}^2$
for BMLJ and WAHN at the lowest equilibrated temperatures. In the case
of BMLJ, in fact, we find a bump at large values in the distribution
of ${E_i^{u}}^2$ for small particles. We also often observed
correlated, string-like rearrangements of large particles in the
unstable modes of BMLJ. This feature is exemplified in the snapshots of
Fig.~\ref{fig:modes2}, where we show the 
unstable modes of a typical quasisaddle sampled in deeply supercooled
BMLJ. By comparison, unstable modes in WAHN tend to involve larger
and more compact clusters of particles and to possess a more
pronounced spatial overlap. These features, in the light of our previous
investigations~\cite{coslovich06}, should influence the
dynamical processes within the $\beta$-relaxation timescale, and could
provide the basis for understanding the microscopic origin of
dynamical heterogeneities~\cite{widmercooper04} on longer timescales. Analysis of the
connectivity between stationary points could also help explaining the
relative weight of different dynamical processes ---string-like
motions~\cite{donati99,schroeder}, democratic 
rearrangements~\cite{appignanesi}--- observed in supercooled
Lennard-Jones mixtures.
More detailed studies along this direction will require significant
additional work.

\section{Conclusions}

Molecular Dynamics simulations of supercooled Lennard-Jones mixtures
continue to provide a useful benchmark for theories and paradigms of
the glass-transition. A minimal exploration of the field of parameters
of the Lennard-Jones potential for binary mixtures has revealed a 
rich phenomenology. In particular, we found a systematic variation
of fragility, i.e., a varying degree of super-Arrhenius behavior of
dynamical properties. By analyzing the temperature dependence of
effective activation energies for relaxation, we found that fragility
increases by increasing size ratio $\lambda=\sigma_{22}/\sigma_{11}$
in equimolar, additive mixtures. Two prototypical mixtures, the
BMLJ mixture of Kob and Andersen~\cite{ka1} and the one of
Wahnstr\"om~\cite{wahnstrom}, also have different fragility
indexes. As an interesting result, we also found that fragility does not
change appreciably with pressure in BMLJ.

In a previous paper~\cite{coslovich07a}, we discussed these trends in terms of the thermal
rate of growth of locally preferred structures upon supercooling. Here, we
have investigated the different, complementary role of the
potential energy surface explored in 
the supercooled regime. We have adopted a simple, non-topographic
approach and analyzed some statistical properties of the PES, with
particular focus on high-order stationary points. We have provided an
estimate of average potential energy barriers and found
a striking correlation with the fragility of the mixture: the more
fragile the mixture, the more 
pronounced the increase of energy barriers upon
supercooling. Not ignoring the role of metabasins~\cite{doliwa03} and multistep
processes~\cite{middleton01b}, an increase of energy barriers between
single saddles appears to be a simple, possible mechanism for super-Arrhenius
behavior of dynamical properties in fragile glass-formers. 
We have also found that a proper characterization of the saddles'
density of states will already encode the relevant information about
fragility. On the other hand, mean frequencies of stable and unstable modes
do not provide robust correlations to fragility. Their strong
variations with density along different isobars in BMLJ, in fact, are
not accompanied by a significant change in fragility.

As a general rule, unstable modes of saddles become more and more
localized upon supercooling, but this feature is sharper and 
more pronounced, the more fragile is the mixture. This can be 
interpreted as the counterpart of a more rapid growth, upon supercooling, of slow
domains, characterized by distinct locally preferred
structures~\cite{coslovich07a}. From some preliminary 
calculations, we have found, as expected, that particles at the center of
locally preferred structures are stabilized and are not involved in
the unstable modes. Thus, the study of the potential
energy surface presented in this work and our previous microstructural
analysis~\cite{coslovich07a} complement each other very well. Formation of stable,
long-lived structures, such as icosahedral domains in additive
mixtures, could correspond to deeper traps in the energy landscape, thus forcing
relaxation over larger energy barriers. On the other hand, frustration
of stable prismatic domains could be the origin of the limited growth
of potential energy barriers in BMLJ. The two approaches together may thus provide an
 intriguing picture of the fragile vs. strong behavior of
glass-former, bridging the ideas of frustration in supercooled
liquids~\cite{tarjus05} and roughness of the energy
landscape~\cite{stillinger95}.   
Assessment of such picture by studying different interactions remains
an open problem for further investigations.

\begin{acknowledgments}
The authors would like to thank A.~Cavagna for useful discussions and a
critical reading of the manuscript.
Computational resources for the present work have been partly obtained
through a grant from  ``Iniziativa Trasversale di Calcolo Parallelo'' of
the Italian {\em CNR-Istituto Nazionale per la Fisica della Materia}
(CNR-INFM) and partly within the agreement between the University of
Trieste and the Consorzio Interuniversitario CINECA (Italy).
\end{acknowledgments}

\bibliographystyle{apsrev}
\bibliography{references}

\begin{thebibliography}{65}
\expandafter\ifx\csname natexlab\endcsname\relax\def\natexlab#1{#1}\fi
\expandafter\ifx\csname bibnamefont\endcsname\relax
  \def\bibnamefont#1{#1}\fi
\expandafter\ifx\csname bibfnamefont\endcsname\relax
  \def\bibfnamefont#1{#1}\fi
\expandafter\ifx\csname citenamefont\endcsname\relax
  \def\citenamefont#1{#1}\fi
\expandafter\ifx\csname url\endcsname\relax
  \def\url#1{\texttt{#1}}\fi
\expandafter\ifx\csname urlprefix\endcsname\relax\def\urlprefix{URL }\fi
\providecommand{\bibinfo}[2]{#2}
\providecommand{\eprint}[2][]{\url{#2}}

\bibitem[{\citenamefont{Sciortino}(2005)}]{sciortino05}
\bibinfo{author}{\bibfnamefont{F.}~\bibnamefont{Sciortino}},
  \bibinfo{journal}{J. Stat. Mech.: Theory Exp.} p. \bibinfo{pages}{P05015}
  (\bibinfo{year}{2005}).

\bibitem[{\citenamefont{Wales}(2003)}]{book:wales}
\bibinfo{author}{\bibfnamefont{D.~J.} \bibnamefont{Wales}},
  \emph{\bibinfo{title}{Energy Landscapes}} (\bibinfo{publisher}{Cambridge
  University Press}, \bibinfo{address}{Cambridge}, \bibinfo{year}{2003}).

\bibitem[{\citenamefont{Stillinger and Weber}(1982)}]{sw82}
\bibinfo{author}{\bibfnamefont{F.~H.} \bibnamefont{Stillinger}}
  \bibnamefont{and} \bibinfo{author}{\bibfnamefont{T.~A.} \bibnamefont{Weber}},
  \bibinfo{journal}{Phys. Rev A} \textbf{\bibinfo{volume}{25}},
  \bibinfo{pages}{983} (\bibinfo{year}{1982}).

\bibitem[{\citenamefont{Sastry et~al.}(1998)\citenamefont{Sastry, Debenedetti,
  and Stillinger}}]{sastry98}
\bibinfo{author}{\bibfnamefont{S.}~\bibnamefont{Sastry}},
  \bibinfo{author}{\bibfnamefont{P.~G.} \bibnamefont{Debenedetti}},
  \bibnamefont{and} \bibinfo{author}{\bibfnamefont{F.~H.}
  \bibnamefont{Stillinger}}, \bibinfo{journal}{Nature}
  \textbf{\bibinfo{volume}{393}}, \bibinfo{pages}{554} (\bibinfo{year}{1998}).

\bibitem[{\citenamefont{Stillinger}(1990)}]{stillinger}
\bibinfo{author}{\bibfnamefont{F.~H.} \bibnamefont{Stillinger}},
  \bibinfo{journal}{Phys. Rev. B} \textbf{\bibinfo{volume}{41}},
  \bibinfo{pages}{2409} (\bibinfo{year}{1990}).

\bibitem[{\citenamefont{Doliwa and Heuer}(2003{\natexlab{a}})}]{doliwa03}
\bibinfo{author}{\bibfnamefont{B.}~\bibnamefont{Doliwa}} \bibnamefont{and}
  \bibinfo{author}{\bibfnamefont{A.}~\bibnamefont{Heuer}},
  \bibinfo{journal}{Phys. Rev. Lett.} \textbf{\bibinfo{volume}{91}},
  \bibinfo{pages}{235501} (\bibinfo{year}{2003}{\natexlab{a}}).

\bibitem[{\citenamefont{Doliwa and Heuer}(2003{\natexlab{b}})}]{doliwa03a}
\bibinfo{author}{\bibfnamefont{B.}~\bibnamefont{Doliwa}} \bibnamefont{and}
  \bibinfo{author}{\bibfnamefont{A.}~\bibnamefont{Heuer}},
  \bibinfo{journal}{Phys. Rev. E} \textbf{\bibinfo{volume}{67}},
  \bibinfo{pages}{031506} (\bibinfo{year}{2003}{\natexlab{b}}).

\bibitem[{\citenamefont{Doliwa and Heuer}(2003{\natexlab{c}})}]{doliwa03b}
\bibinfo{author}{\bibfnamefont{B.}~\bibnamefont{Doliwa}} \bibnamefont{and}
  \bibinfo{author}{\bibfnamefont{A.}~\bibnamefont{Heuer}},
  \bibinfo{journal}{Phys. Rev. E} \textbf{\bibinfo{volume}{67}},
  \bibinfo{pages}{030501} (\bibinfo{year}{2003}{\natexlab{c}}).

\bibitem[{\citenamefont{Denny et~al.}(2003)\citenamefont{Denny, Reichman, and
  Bouchaud}}]{denny03}
\bibinfo{author}{\bibfnamefont{R.~A.} \bibnamefont{Denny}},
  \bibinfo{author}{\bibfnamefont{D.~R.} \bibnamefont{Reichman}},
  \bibnamefont{and} \bibinfo{author}{\bibfnamefont{J.-P.}
  \bibnamefont{Bouchaud}}, \bibinfo{journal}{Phys. Rev. Lett.}
  \textbf{\bibinfo{volume}{90}}, \bibinfo{pages}{025503}
  (\bibinfo{year}{2003}).

\bibitem[{\citenamefont{Stillinger}(1995)}]{stillinger95}
\bibinfo{author}{\bibfnamefont{F.~H.} \bibnamefont{Stillinger}},
  \bibinfo{journal}{Science} \textbf{\bibinfo{volume}{267}},
  \bibinfo{pages}{1935} (\bibinfo{year}{1995}).

\bibitem[{\citenamefont{Sastry}(2001)}]{sastry01}
\bibinfo{author}{\bibfnamefont{S.}~\bibnamefont{Sastry}},
  \bibinfo{journal}{Nature} \textbf{\bibinfo{volume}{409}},
  \bibinfo{pages}{164} (\bibinfo{year}{2001}).

\bibitem[{\citenamefont{Speedy}(1999)}]{speedy99}
\bibinfo{author}{\bibfnamefont{R.~J.} \bibnamefont{Speedy}},
  \bibinfo{journal}{J. Phys. Chem. B} \textbf{\bibinfo{volume}{103}},
  \bibinfo{pages}{4060} (\bibinfo{year}{1999}).

\bibitem[{\citenamefont{Ruocco et~al.}(2004)\citenamefont{Ruocco, Sciortino,
  Zamponi, {De Michele}, and Scopigno}}]{ruocco04}
\bibinfo{author}{\bibfnamefont{G.}~\bibnamefont{Ruocco}},
  \bibinfo{author}{\bibfnamefont{F.}~\bibnamefont{Sciortino}},
  \bibinfo{author}{\bibfnamefont{F.}~\bibnamefont{Zamponi}},
  \bibinfo{author}{\bibfnamefont{C.}~\bibnamefont{{De Michele}}},
  \bibnamefont{and} \bibinfo{author}{\bibfnamefont{T.}~\bibnamefont{Scopigno}},
  \bibinfo{journal}{J. Chem. Phys.} \textbf{\bibinfo{volume}{120}},
  \bibinfo{pages}{10666} (\bibinfo{year}{2004}).

\bibitem[{\citenamefont{Tarjus et~al.}(2004)\citenamefont{Tarjus, Kivelson,
  Mossa, and Alba-Simionesco}}]{tarjus04}
\bibinfo{author}{\bibfnamefont{G.}~\bibnamefont{Tarjus}},
  \bibinfo{author}{\bibfnamefont{D.}~\bibnamefont{Kivelson}},
  \bibinfo{author}{\bibfnamefont{S.}~\bibnamefont{Mossa}}, \bibnamefont{and}
  \bibinfo{author}{\bibfnamefont{C.}~\bibnamefont{Alba-Simionesco}},
  \bibinfo{journal}{J. Chem. Phys.} \textbf{\bibinfo{volume}{120}},
  \bibinfo{pages}{6135} (\bibinfo{year}{2004}).

\bibitem[{\citenamefont{Middleton and Wales}(2001)}]{middleton01b}
\bibinfo{author}{\bibfnamefont{T.~F.} \bibnamefont{Middleton}}
  \bibnamefont{and} \bibinfo{author}{\bibfnamefont{D.~J.} \bibnamefont{Wales}},
  \bibinfo{journal}{Phys. Rev. B} \textbf{\bibinfo{volume}{64}},
  \bibinfo{pages}{024205} (\bibinfo{year}{2001}).

\bibitem[{\citenamefont{Middleton and Wales}(2003)}]{middleton03}
\bibinfo{author}{\bibfnamefont{T.~F.} \bibnamefont{Middleton}}
  \bibnamefont{and} \bibinfo{author}{\bibfnamefont{D.~J.} \bibnamefont{Wales}},
  \bibinfo{journal}{J. Chem. Phys.} \textbf{\bibinfo{volume}{118}},
  \bibinfo{pages}{4583} (\bibinfo{year}{2003}).

\bibitem[{\citenamefont{Angelani et~al.}(2003)\citenamefont{Angelani, Ruocco,
  Sampoli, and Sciortino}}]{angelani03}
\bibinfo{author}{\bibfnamefont{L.}~\bibnamefont{Angelani}},
  \bibinfo{author}{\bibfnamefont{G.}~\bibnamefont{Ruocco}},
  \bibinfo{author}{\bibfnamefont{M.}~\bibnamefont{Sampoli}}, \bibnamefont{and}
  \bibinfo{author}{\bibfnamefont{F.}~\bibnamefont{Sciortino}},
  \bibinfo{journal}{J. Chem. Phys.} \textbf{\bibinfo{volume}{119}},
  \bibinfo{pages}{2120} (\bibinfo{year}{2003}).

\bibitem[{\citenamefont{Doye and Wales}(2002)}]{doye02}
\bibinfo{author}{\bibfnamefont{J.~P.~K.} \bibnamefont{Doye}} \bibnamefont{and}
  \bibinfo{author}{\bibfnamefont{D.~J.} \bibnamefont{Wales}},
  \bibinfo{journal}{J. Chem. Phys.} \textbf{\bibinfo{volume}{116}},
  \bibinfo{pages}{3777} (\bibinfo{year}{2002}).

\bibitem[{\citenamefont{Wales and Doye}(2003)}]{wales03}
\bibinfo{author}{\bibfnamefont{D.~J.} \bibnamefont{Wales}} \bibnamefont{and}
  \bibinfo{author}{\bibfnamefont{J.~P.~K.} \bibnamefont{Doye}},
  \bibinfo{journal}{J. Chem. Phys.} \textbf{\bibinfo{volume}{119}},
  \bibinfo{pages}{12409} (\bibinfo{year}{2003}).

\bibitem[{\citenamefont{Sampoli et~al.}(2003)\citenamefont{Sampoli, Benassi,
  Eramo, Angelani, and Ruocco}}]{sampoli03}
\bibinfo{author}{\bibfnamefont{M.}~\bibnamefont{Sampoli}},
  \bibinfo{author}{\bibfnamefont{P.}~\bibnamefont{Benassi}},
  \bibinfo{author}{\bibfnamefont{R.}~\bibnamefont{Eramo}},
  \bibinfo{author}{\bibfnamefont{L.}~\bibnamefont{Angelani}}, \bibnamefont{and}
  \bibinfo{author}{\bibfnamefont{G.}~\bibnamefont{Ruocco}},
  \bibinfo{journal}{J. Phys.: Condens. Matter} \textbf{\bibinfo{volume}{15}},
  \bibinfo{pages}{S1227} (\bibinfo{year}{2003}).

\bibitem[{\citenamefont{Angelani et~al.}(2000)\citenamefont{Angelani, {Di
  Leonardo}, Ruocco, Scala, and Sciortino}}]{angelani00}
\bibinfo{author}{\bibfnamefont{L.}~\bibnamefont{Angelani}},
  \bibinfo{author}{\bibfnamefont{R.}~\bibnamefont{{Di Leonardo}}},
  \bibinfo{author}{\bibfnamefont{G.}~\bibnamefont{Ruocco}},
  \bibinfo{author}{\bibfnamefont{A.}~\bibnamefont{Scala}}, \bibnamefont{and}
  \bibinfo{author}{\bibfnamefont{F.}~\bibnamefont{Sciortino}},
  \bibinfo{journal}{Phys. Rev. Lett.} \textbf{\bibinfo{volume}{8}},
  \bibinfo{pages}{5356} (\bibinfo{year}{2000}).

\bibitem[{\citenamefont{Angelani et~al.}(2002)\citenamefont{Angelani, {Di
  Leonardo}, Ruocco, Scala, and Sciortino}}]{angelani02}
\bibinfo{author}{\bibfnamefont{L.}~\bibnamefont{Angelani}},
  \bibinfo{author}{\bibfnamefont{R.}~\bibnamefont{{Di Leonardo}}},
  \bibinfo{author}{\bibfnamefont{G.}~\bibnamefont{Ruocco}},
  \bibinfo{author}{\bibfnamefont{A.}~\bibnamefont{Scala}}, \bibnamefont{and}
  \bibinfo{author}{\bibfnamefont{F.}~\bibnamefont{Sciortino}},
  \bibinfo{journal}{J. Chem. Phys.} \textbf{\bibinfo{volume}{116}},
  \bibinfo{pages}{10297} (\bibinfo{year}{2002}).

\bibitem[{\citenamefont{Coslovich and Pastore}(2006)}]{coslovich06}
\bibinfo{author}{\bibfnamefont{D.}~\bibnamefont{Coslovich}} \bibnamefont{and}
  \bibinfo{author}{\bibfnamefont{G.}~\bibnamefont{Pastore}},
  \bibinfo{journal}{Europhys. Lett.} \textbf{\bibinfo{volume}{75}},
  \bibinfo{pages}{784} (\bibinfo{year}{2006}).

\bibitem[{\citenamefont{Cavagna}(2001)}]{cavagna01c}
\bibinfo{author}{\bibfnamefont{A.}~\bibnamefont{Cavagna}},
  \bibinfo{journal}{Europhys. Lett.} \textbf{\bibinfo{volume}{53}},
  \bibinfo{pages}{490} (\bibinfo{year}{2001}).

\bibitem[{\citenamefont{Zamponi et~al.}(2003)\citenamefont{Zamponi, Angelani,
  Cugliandolo, Kurchan, and Ruocco}}]{zamponi03}
\bibinfo{author}{\bibfnamefont{F.}~\bibnamefont{Zamponi}},
  \bibinfo{author}{\bibfnamefont{L.}~\bibnamefont{Angelani}},
  \bibinfo{author}{\bibfnamefont{L.~F.} \bibnamefont{Cugliandolo}},
  \bibinfo{author}{\bibfnamefont{J.}~\bibnamefont{Kurchan}}, \bibnamefont{and}
  \bibinfo{author}{\bibfnamefont{G.}~\bibnamefont{Ruocco}},
  \bibinfo{journal}{J. Phys. A: Math. Gen.} \textbf{\bibinfo{volume}{36}},
  \bibinfo{pages}{8565} (\bibinfo{year}{2003}).

\bibitem[{\citenamefont{Keyes et~al.}(2002)\citenamefont{Keyes, Chowdhary, and
  Kim}}]{keyes02b}
\bibinfo{author}{\bibfnamefont{T.}~\bibnamefont{Keyes}},
  \bibinfo{author}{\bibfnamefont{J.}~\bibnamefont{Chowdhary}},
  \bibnamefont{and} \bibinfo{author}{\bibfnamefont{J.}~\bibnamefont{Kim}},
  \bibinfo{journal}{Phys. Rev. E} \textbf{\bibinfo{volume}{66}},
  \bibinfo{pages}{051110} (\bibinfo{year}{2002}).

\bibitem[{\citenamefont{Andronico et~al.}(2004)\citenamefont{Andronico,
  Angelani, Ruocco, and Zamponi}}]{andronico04}
\bibinfo{author}{\bibfnamefont{A.}~\bibnamefont{Andronico}},
  \bibinfo{author}{\bibfnamefont{L.}~\bibnamefont{Angelani}},
  \bibinfo{author}{\bibfnamefont{G.}~\bibnamefont{Ruocco}}, \bibnamefont{and}
  \bibinfo{author}{\bibfnamefont{F.}~\bibnamefont{Zamponi}},
  \bibinfo{journal}{Phys. Rev. E} \textbf{\bibinfo{volume}{70}},
  \bibinfo{pages}{041101} (\bibinfo{year}{2004}).

\bibitem[{\citenamefont{Cavagna et~al.}(2001)\citenamefont{Cavagna, Giardina,
  and Parisi}}]{cavagna01b}
\bibinfo{author}{\bibfnamefont{A.}~\bibnamefont{Cavagna}},
  \bibinfo{author}{\bibfnamefont{I.}~\bibnamefont{Giardina}}, \bibnamefont{and}
  \bibinfo{author}{\bibfnamefont{G.}~\bibnamefont{Parisi}},
  \bibinfo{journal}{J. Phys. A: Math. Gen.} \textbf{\bibinfo{volume}{34}},
  \bibinfo{pages}{5317} (\bibinfo{year}{2001}).

\bibitem[{\citenamefont{Grigera et~al.}(2002)\citenamefont{Grigera, Cavagna,
  Giardina, and Parisi}}]{grigera02}
\bibinfo{author}{\bibfnamefont{T.}~\bibnamefont{Grigera}},
  \bibinfo{author}{\bibfnamefont{A.}~\bibnamefont{Cavagna}},
  \bibinfo{author}{\bibfnamefont{I.}~\bibnamefont{Giardina}}, \bibnamefont{and}
  \bibinfo{author}{\bibfnamefont{G.}~\bibnamefont{Parisi}},
  \bibinfo{journal}{Phys. Rev. Lett.} \textbf{\bibinfo{volume}{88}},
  \bibinfo{pages}{055502} (\bibinfo{year}{2002}).

\bibitem[{\citenamefont{Shah and Chakravarty}(2001)}]{shah01}
\bibinfo{author}{\bibfnamefont{P.}~\bibnamefont{Shah}} \bibnamefont{and}
  \bibinfo{author}{\bibfnamefont{C.}~\bibnamefont{Chakravarty}},
  \bibinfo{journal}{J. Chem. Physics} \textbf{\bibinfo{volume}{115}},
  \bibinfo{pages}{8784} (\bibinfo{year}{2001}).

\bibitem[{\citenamefont{Chakraborty and Chakravarty}(2006)}]{chakraborty06}
\bibinfo{author}{\bibfnamefont{S.~N.} \bibnamefont{Chakraborty}}
  \bibnamefont{and}
  \bibinfo{author}{\bibfnamefont{C.}~\bibnamefont{Chakravarty}},
  \bibinfo{journal}{J. Chem. Phys.} \textbf{\bibinfo{volume}{124}},
  \bibinfo{pages}{014507} (\bibinfo{year}{2006}).

\bibitem[{\citenamefont{Angelani et~al.}(2004)\citenamefont{Angelani, {De
  Michele}, Ruocco, and Sciortino}}]{angelani04}
\bibinfo{author}{\bibfnamefont{L.}~\bibnamefont{Angelani}},
  \bibinfo{author}{\bibfnamefont{C.}~\bibnamefont{{De Michele}}},
  \bibinfo{author}{\bibfnamefont{G.}~\bibnamefont{Ruocco}}, \bibnamefont{and}
  \bibinfo{author}{\bibfnamefont{F.}~\bibnamefont{Sciortino}},
  \bibinfo{journal}{J. Chem. Phys.} \textbf{\bibinfo{volume}{121}},
  \bibinfo{pages}{7533} (\bibinfo{year}{2004}).

\bibitem[{\citenamefont{{De Michele} et~al.}(2004)\citenamefont{{De Michele},
  Sciortino, and Coniglio}}]{demichele04}
\bibinfo{author}{\bibfnamefont{C.}~\bibnamefont{{De Michele}}},
  \bibinfo{author}{\bibfnamefont{F.}~\bibnamefont{Sciortino}},
  \bibnamefont{and} \bibinfo{author}{\bibfnamefont{A.}~\bibnamefont{Coniglio}},
  \bibinfo{journal}{J. Phys.: Condens. Matter} \textbf{\bibinfo{volume}{16}},
  \bibinfo{pages}{L489} (\bibinfo{year}{2004}).

\bibitem[{\citenamefont{Coslovich and Pastore}(2007)}]{coslovich07a}
\bibinfo{author}{\bibfnamefont{D.}~\bibnamefont{Coslovich}} \bibnamefont{and}
  \bibinfo{author}{\bibfnamefont{G.}~\bibnamefont{Pastore}},
  \bibinfo{journal}{J. Chem. Phys.} \textbf{\bibinfo{volume}{127}},
  \bibinfo{pages}{124504} (\bibinfo{year}{2007}).

\bibitem[{\citenamefont{Kob and Andersen}(1995{\natexlab{a}})}]{ka1}
\bibinfo{author}{\bibfnamefont{W.}~\bibnamefont{Kob}} \bibnamefont{and}
  \bibinfo{author}{\bibfnamefont{H.~C.} \bibnamefont{Andersen}},
  \bibinfo{journal}{Phys. Rev. E} \textbf{\bibinfo{volume}{51}},
  \bibinfo{pages}{4626} (\bibinfo{year}{1995}{\natexlab{a}}).

\bibitem[{\citenamefont{Kob and Andersen}(1995{\natexlab{b}})}]{ka2}
\bibinfo{author}{\bibfnamefont{W.}~\bibnamefont{Kob}} \bibnamefont{and}
  \bibinfo{author}{\bibfnamefont{H.~C.} \bibnamefont{Andersen}},
  \bibinfo{journal}{Phys. Rev. E} \textbf{\bibinfo{volume}{52}},
  \bibinfo{pages}{4134} (\bibinfo{year}{1995}{\natexlab{b}}).

\bibitem[{\citenamefont{Stoddard and Ford}(1973)}]{stoddard73}
\bibinfo{author}{\bibfnamefont{S.~D.} \bibnamefont{Stoddard}} \bibnamefont{and}
  \bibinfo{author}{\bibfnamefont{J.}~\bibnamefont{Ford}},
  \bibinfo{journal}{Phys. Rev. A} \textbf{\bibinfo{volume}{8}},
  \bibinfo{pages}{1504} (\bibinfo{year}{1973}).

\bibitem[{\citenamefont{Shah and Chakravarty}(2003)}]{shah03}
\bibinfo{author}{\bibfnamefont{P.}~\bibnamefont{Shah}} \bibnamefont{and}
  \bibinfo{author}{\bibfnamefont{C.}~\bibnamefont{Chakravarty}},
  \bibinfo{journal}{J. Chem. Phys.} \textbf{\bibinfo{volume}{118}},
  \bibinfo{pages}{2342} (\bibinfo{year}{2003}).

\bibitem[{\citenamefont{Wahnstr{\"o}m}(1991)}]{wahnstrom}
\bibinfo{author}{\bibfnamefont{G.}~\bibnamefont{Wahnstr{\"o}m}},
  \bibinfo{journal}{Phys. Rev. A} \textbf{\bibinfo{volume}{44}},
  \bibinfo{pages}{3752} (\bibinfo{year}{1991}).

\bibitem[{\citenamefont{Bond et~al.}(1999)\citenamefont{Bond, Leimkuhler, and
  Laird}}]{bond}
\bibinfo{author}{\bibfnamefont{S.~D.} \bibnamefont{Bond}},
  \bibinfo{author}{\bibfnamefont{B.~J.} \bibnamefont{Leimkuhler}},
  \bibnamefont{and} \bibinfo{author}{\bibfnamefont{B.~B.} \bibnamefont{Laird}},
  \bibinfo{journal}{J. Comp. Phys.} \textbf{\bibinfo{volume}{151}},
  \bibinfo{pages}{114} (\bibinfo{year}{1999}).

\bibitem[{\citenamefont{Nos\'e}(2001)}]{nose01}
\bibinfo{author}{\bibfnamefont{S.}~\bibnamefont{Nos\'e}}, \bibinfo{journal}{J.
  Phys. Soc. Jap.} \textbf{\bibinfo{volume}{70}}, \bibinfo{pages}{75}
  (\bibinfo{year}{2001}).

\bibitem[{\citenamefont{Allen and Tildesley}(1987)}]{book:at}
\bibinfo{author}{\bibfnamefont{M.~P.} \bibnamefont{Allen}} \bibnamefont{and}
  \bibinfo{author}{\bibfnamefont{D.~J.} \bibnamefont{Tildesley}},
  \emph{\bibinfo{title}{Computer Simulation of Liquids}}
  (\bibinfo{publisher}{Clarendon Press}, \bibinfo{year}{1987}).

\bibitem[{\citenamefont{Liu and Nocedal}(1989)}]{nocedal}
\bibinfo{author}{\bibfnamefont{D.~C.} \bibnamefont{Liu}} \bibnamefont{and}
  \bibinfo{author}{\bibfnamefont{J.}~\bibnamefont{Nocedal}},
  \bibinfo{journal}{Math. Program.} \textbf{\bibinfo{volume}{45}},
  \bibinfo{pages}{503} (\bibinfo{year}{1989}).

\bibitem[{\citenamefont{Grigera}(2006)}]{grigera06}
\bibinfo{author}{\bibfnamefont{T.~S.} \bibnamefont{Grigera}},
  \bibinfo{journal}{J. Chem. Phys.} \textbf{\bibinfo{volume}{124}},
  \bibinfo{pages}{064502} (\bibinfo{year}{2006}).

\bibitem[{\citenamefont{Kivelson et~al.}(1996)\citenamefont{Kivelson, Tarjus,
  Zhao, and Kivelson}}]{kivelson96}
\bibinfo{author}{\bibfnamefont{D.}~\bibnamefont{Kivelson}},
  \bibinfo{author}{\bibfnamefont{G.}~\bibnamefont{Tarjus}},
  \bibinfo{author}{\bibfnamefont{X.}~\bibnamefont{Zhao}}, \bibnamefont{and}
  \bibinfo{author}{\bibfnamefont{S.~A.} \bibnamefont{Kivelson}},
  \bibinfo{journal}{Phys. Rev. E} \textbf{\bibinfo{volume}{53}},
  \bibinfo{pages}{751} (\bibinfo{year}{1996}).

\bibitem[{\citenamefont{Ferrer et~al.}(1999)\citenamefont{Ferrer, Sakai,
  Kivelson, and Alba-Simionesco}}]{ferrer99}
\bibinfo{author}{\bibfnamefont{M.~L.} \bibnamefont{Ferrer}},
  \bibinfo{author}{\bibfnamefont{H.}~\bibnamefont{Sakai}},
  \bibinfo{author}{\bibfnamefont{D.}~\bibnamefont{Kivelson}}, \bibnamefont{and}
  \bibinfo{author}{\bibfnamefont{C.}~\bibnamefont{Alba-Simionesco}},
  \bibinfo{journal}{J. Phys. Chem. B} \textbf{\bibinfo{volume}{103}},
  \bibinfo{pages}{4191} (\bibinfo{year}{1999}).

\bibitem[{\citenamefont{Tarjus et~al.}(2000)\citenamefont{Tarjus, Kivelson, and
  Viot}}]{tarjus00}
\bibinfo{author}{\bibfnamefont{G.}~\bibnamefont{Tarjus}},
  \bibinfo{author}{\bibfnamefont{D.}~\bibnamefont{Kivelson}}, \bibnamefont{and}
  \bibinfo{author}{\bibfnamefont{P.}~\bibnamefont{Viot}}, \bibinfo{journal}{J.
  Phys.: Condens. Matter} \textbf{\bibinfo{volume}{12}}, \bibinfo{pages}{6497}
  (\bibinfo{year}{2000}).

\bibitem[{\citenamefont{Novikov et~al.}(2005)\citenamefont{Novikov, Ding, and
  Sokolov}}]{novikov05}
\bibinfo{author}{\bibfnamefont{V.~N.} \bibnamefont{Novikov}},
  \bibinfo{author}{\bibfnamefont{Y.}~\bibnamefont{Ding}}, \bibnamefont{and}
  \bibinfo{author}{\bibfnamefont{A.~P.} \bibnamefont{Sokolov}},
  \bibinfo{journal}{Phys. Rev. E} \textbf{\bibinfo{volume}{71}},
  \bibinfo{pages}{061501} (\bibinfo{year}{2005}).

\bibitem[{\citenamefont{Angell et~al.}(1999)\citenamefont{Angell, Richards, and
  Velikov}}]{angell99}
\bibinfo{author}{\bibfnamefont{C.~A.} \bibnamefont{Angell}},
  \bibinfo{author}{\bibfnamefont{B.~E.} \bibnamefont{Richards}},
  \bibnamefont{and} \bibinfo{author}{\bibfnamefont{V.}~\bibnamefont{Velikov}},
  \bibinfo{journal}{J. Phys.: Condens. Matter} \textbf{\bibinfo{volume}{11}},
  \bibinfo{pages}{A75} (\bibinfo{year}{1999}).

\bibitem[{\citenamefont{Bembenek and Laird}(1996)}]{bembenek}
\bibinfo{author}{\bibfnamefont{S.~D.} \bibnamefont{Bembenek}} \bibnamefont{and}
  \bibinfo{author}{\bibfnamefont{B.~B.} \bibnamefont{Laird}},
  \bibinfo{journal}{J. Chem. Phys.} \textbf{\bibinfo{volume}{104}},
  \bibinfo{pages}{5199} (\bibinfo{year}{1996}).

\bibitem[{\citenamefont{Gezelter et~al.}(1997)\citenamefont{Gezelter, Rabani,
  and Berne}}]{gezelter}
\bibinfo{author}{\bibfnamefont{J.~D.} \bibnamefont{Gezelter}},
  \bibinfo{author}{\bibfnamefont{E.}~\bibnamefont{Rabani}}, \bibnamefont{and}
  \bibinfo{author}{\bibfnamefont{B.~J.} \bibnamefont{Berne}},
  \bibinfo{journal}{J. Chem. Phys.} \textbf{\bibinfo{volume}{107}},
  \bibinfo{pages}{4618} (\bibinfo{year}{1997}).

\bibitem[{\citenamefont{Keyes et~al.}(1997)\citenamefont{Keyes, Vijayadamodar,
  and Zurcher}}]{keyes97}
\bibinfo{author}{\bibfnamefont{T.}~\bibnamefont{Keyes}},
  \bibinfo{author}{\bibfnamefont{G.~V.} \bibnamefont{Vijayadamodar}},
  \bibnamefont{and} \bibinfo{author}{\bibfnamefont{U.}~\bibnamefont{Zurcher}},
  \bibinfo{journal}{J. Chem. Phys.} \textbf{\bibinfo{volume}{106}},
  \bibinfo{pages}{4651} (\bibinfo{year}{1997}).

\bibitem[{\citenamefont{Broderix et~al.}(2000)\citenamefont{Broderix,
  Bhattacharya, Cavagna, Zippelius, and Giardina}}]{broderix00}
\bibinfo{author}{\bibfnamefont{K.}~\bibnamefont{Broderix}},
  \bibinfo{author}{\bibfnamefont{K.~K.} \bibnamefont{Bhattacharya}},
  \bibinfo{author}{\bibfnamefont{A.}~\bibnamefont{Cavagna}},
  \bibinfo{author}{\bibfnamefont{A.}~\bibnamefont{Zippelius}},
  \bibnamefont{and} \bibinfo{author}{\bibfnamefont{I.}~\bibnamefont{Giardina}},
  \bibinfo{journal}{Phys. Rev. Lett.} \textbf{\bibinfo{volume}{85}},
  \bibinfo{pages}{5360} (\bibinfo{year}{2000}).

\bibitem[{\citenamefont{Cavagna et~al.}(2003)\citenamefont{Cavagna, Giardina,
  and Grigera}}]{cavagna03}
\bibinfo{author}{\bibfnamefont{A.}~\bibnamefont{Cavagna}},
  \bibinfo{author}{\bibfnamefont{I.}~\bibnamefont{Giardina}}, \bibnamefont{and}
  \bibinfo{author}{\bibfnamefont{T.}~\bibnamefont{Grigera}},
  \bibinfo{journal}{J. Phys. A: Math. Gen.} \textbf{\bibinfo{volume}{36}},
  \bibinfo{pages}{10721} (\bibinfo{year}{2003}).

\bibitem[{\citenamefont{Keyes}(1994)}]{keyes95a}
\bibinfo{author}{\bibfnamefont{T.}~\bibnamefont{Keyes}}, \bibinfo{journal}{J.
  Chem. Phys.} \textbf{\bibinfo{volume}{101}}, \bibinfo{pages}{5081}
  (\bibinfo{year}{1994}).

\bibitem[{\citenamefont{Vijayadamodar and Nitzan}(1995)}]{nitzan95}
\bibinfo{author}{\bibfnamefont{G.}~\bibnamefont{Vijayadamodar}}
  \bibnamefont{and} \bibinfo{author}{\bibfnamefont{A.}~\bibnamefont{Nitzan}},
  \bibinfo{journal}{J. Chem. Phys.} \textbf{\bibinfo{volume}{103}},
  \bibinfo{pages}{2169} (\bibinfo{year}{1995}).

\bibitem[{\citenamefont{Li and Keyes}(1999)}]{li99}
\bibinfo{author}{\bibfnamefont{W.-X.} \bibnamefont{Li}} \bibnamefont{and}
  \bibinfo{author}{\bibfnamefont{T.}~\bibnamefont{Keyes}}, \bibinfo{journal}{J.
  Chem. Phys.} \textbf{\bibinfo{volume}{111}}, \bibinfo{pages}{5503}
  (\bibinfo{year}{1999}).

\bibitem[{\citenamefont{Berthier and Garrahan}(2003)}]{berthier03}
\bibinfo{author}{\bibfnamefont{L.}~\bibnamefont{Berthier}} \bibnamefont{and}
  \bibinfo{author}{\bibfnamefont{J.~P.} \bibnamefont{Garrahan}},
  \bibinfo{journal}{Phys. Rev. E} \textbf{\bibinfo{volume}{68}},
  \bibinfo{pages}{041201} (\bibinfo{year}{2003}).

\bibitem[{\citenamefont{Jagla}(2001)}]{jagla01}
\bibinfo{author}{\bibfnamefont{E.~A.} \bibnamefont{Jagla}},
  \bibinfo{journal}{Mol. Phys.} \textbf{\bibinfo{volume}{99}},
  \bibinfo{pages}{753} (\bibinfo{year}{2001}).

\bibitem[{\citenamefont{Widmer-Cooper et~al.}(2004)\citenamefont{Widmer-Cooper,
  Harrowell, and Fynewever}}]{widmercooper04}
\bibinfo{author}{\bibfnamefont{A.}~\bibnamefont{Widmer-Cooper}},
  \bibinfo{author}{\bibfnamefont{P.}~\bibnamefont{Harrowell}},
  \bibnamefont{and}
  \bibinfo{author}{\bibfnamefont{H.}~\bibnamefont{Fynewever}},
  \bibinfo{journal}{Phys. Rev. Lett.} \textbf{\bibinfo{volume}{93}},
  \bibinfo{pages}{135701} (\bibinfo{year}{2004}).

\bibitem[{\citenamefont{Donati et~al.}(1999)\citenamefont{Donati, Glotzer,
  Poole, Plimpton, and Kob}}]{donati99}
\bibinfo{author}{\bibfnamefont{C.}~\bibnamefont{Donati}},
  \bibinfo{author}{\bibfnamefont{S.~C.} \bibnamefont{Glotzer}},
  \bibinfo{author}{\bibfnamefont{P.~H.} \bibnamefont{Poole}},
  \bibinfo{author}{\bibfnamefont{S.~J.} \bibnamefont{Plimpton}},
  \bibnamefont{and} \bibinfo{author}{\bibfnamefont{W.}~\bibnamefont{Kob}},
  \bibinfo{journal}{Phys. Rev. E} \textbf{\bibinfo{volume}{60}},
  \bibinfo{pages}{3107} (\bibinfo{year}{1999}).

\bibitem[{\citenamefont{Schroeder et~al.}(2000)\citenamefont{Schroeder, Sastry,
  Dyre, and Glotzer}}]{schroeder}
\bibinfo{author}{\bibfnamefont{T.~B.} \bibnamefont{Schroeder}},
  \bibinfo{author}{\bibfnamefont{S.}~\bibnamefont{Sastry}},
  \bibinfo{author}{\bibfnamefont{J.~C.} \bibnamefont{Dyre}}, \bibnamefont{and}
  \bibinfo{author}{\bibfnamefont{S.~C.} \bibnamefont{Glotzer}},
  \bibinfo{journal}{J. Chem. Phys.} \textbf{\bibinfo{volume}{112}},
  \bibinfo{pages}{9834} (\bibinfo{year}{2000}).

\bibitem[{\citenamefont{Appignanesi et~al.}(2006)\citenamefont{Appignanesi,
  {Rodriguez Fris}, Montani, and Kob}}]{appignanesi}
\bibinfo{author}{\bibfnamefont{G.~A.} \bibnamefont{Appignanesi}},
  \bibinfo{author}{\bibfnamefont{J.~A.} \bibnamefont{{Rodriguez Fris}}},
  \bibinfo{author}{\bibfnamefont{R.~A.} \bibnamefont{Montani}},
  \bibnamefont{and} \bibinfo{author}{\bibfnamefont{W.}~\bibnamefont{Kob}},
  \bibinfo{journal}{Phys. Rev. Lett.} \textbf{\bibinfo{volume}{96}},
  \bibinfo{pages}{057801} (\bibinfo{year}{2006}).

\bibitem[{\citenamefont{Tarjus et~al.}(2005)\citenamefont{Tarjus, Kivelson,
  Nussinov, and Viot}}]{tarjus05}
\bibinfo{author}{\bibfnamefont{G.}~\bibnamefont{Tarjus}},
  \bibinfo{author}{\bibfnamefont{S.~A.} \bibnamefont{Kivelson}},
  \bibinfo{author}{\bibfnamefont{Z.}~\bibnamefont{Nussinov}}, \bibnamefont{and}
  \bibinfo{author}{\bibfnamefont{P.}~\bibnamefont{Viot}}, \bibinfo{journal}{J.
  Phys.: Condens. Matter} \textbf{\bibinfo{volume}{17}}, \bibinfo{pages}{R1182}
  (\bibinfo{year}{2005}).

\bibitem[{\citenamefont{Chowdhary and Keyes}(2002)}]{chowdhary02}
\bibinfo{author}{\bibfnamefont{J.}~\bibnamefont{Chowdhary}} \bibnamefont{and}
  \bibinfo{author}{\bibfnamefont{T.}~\bibnamefont{Keyes}},
  \bibinfo{journal}{Physica A} \textbf{\bibinfo{volume}{314}},
  \bibinfo{pages}{575} (\bibinfo{year}{2002}).

\end{thebibliography}

\end{document}